\documentclass[12pt]{article} 
\usepackage{latexsym}
\usepackage{amsmath}
\usepackage{pstricks}
\usepackage{indentfirst}
\newcommand{\ad}{^\dagger }
\newcommand{\becs}{\begin{cases}}
\newcommand{\bem}{\begin{matrix}}
\newcommand{\besp}{\begin{split}} 
 
\newcommand{\Blp}{\Bigl(} 
\newcommand{\Brp}{\Bigr)}

\newcommand{\dg}{^\circ } 
\newcommand{\encs}{\end{cases}}
\newcommand{\enm}{\end{matrix}}
\newcommand{\ensp}{\end{split}}

\newcommand{\lgl}{\langle }

\newcommand{\ot}{\otimes }

\newcommand{\ra}{\rightarrow }

\newcommand{\rgl}{\rangle }
\newcommand{\Rn}{{\rm Rn}}

\newcommand{\tm}{\times }
\newcommand{\Tr}{{\rm Tr}}
\newcommand{\trp}{^\textup{T}}
\newcommand{\vb}{\,|\,}

\newcommand{\bld}[1]{\boldsymbol{#1}}

\newcommand{\scriptl}[1] {{\cal #1}}

\newcommand{\AS}{\scriptl A }

\newcommand{\CS}{\scriptl C }

\newcommand{\HS}{\scriptl H }

\newcommand{\LS}{\scriptl L }

\newcommand{\VS}{\scriptl V }
\newcommand{\WS}{\scriptl W }

\newcommand{\al}{\alpha }
\newcommand{\bt}{\beta }
\newcommand{\gm}{\gamma }
\newcommand{\Gm}{\Gamma }
\newcommand{\dl}{\delta }

\newcommand{\kp}{\kappa }
\newcommand{\lm}{\lambda }

\newcommand{\sg}{\sigma }

\newcommand{\om}{\omega }
\newcommand{\Om}{\Omega }

\newcommand{\lwd}{0.035} 
\newcommand{\lwn}{0.02}  
\newcommand{\rdot}{0.1}
\newcommand{\rodot}{0.1}
\newcommand{\robj}{0.4}
\newcommand{\rbobj}{0.6}
\newcommand{\rfrm}{0.8}
\newcommand{\cldot}{\pscircle*(0,0){\rdot}} 
\newcommand{\opdot}{%
\pscircle[fillcolor=white,fillstyle=solid](0,0){\rodot}} 
\newcommand{\obc}
{\pscircle[fillcolor=white,fillstyle=solid,linewidth=\lwn](0,0){\robj}}
\newcommand{\obx}{
\psframe[fillcolor=white,fillstyle=solid,linewidth=\lwn]%
(-\robj,-\robj)(\robj,\robj)}
\newcommand{\bobx}{
\psframe[fillcolor=white,fillstyle=solid,linewidth=\lwn]%
(-\rbobj,-\rbobj)(\rbobj,\rbobj)}
\newcommand{\obr}[1]{
\psframe[framearc=\rfrm,fillcolor=white,fillstyle=solid,linewidth=\lwn]%
(-#1,-\robj)(#1,\robj)}
\newcommand{\vvv}[1]{\vrule height #1 cm depth #1 cm width 0pt}

\begin{document}

\title{Atemporal Diagrams for Quantum Circuits}


\author{Robert B. Griffiths$^1$, 
Shengjun Wu$^1$, Li Yu$^1$, and Scott M. Cohen$^{1,2}$\\
$^1$Physics Department,
Carnegie-Mellon University\\
Pittsburgh, PA 15213, USA\\
$^2$Physics Department,
Duquesne University,
Pittsburgh, PA 15282, USA}


\date{Version of 23 August 2005}

\maketitle  

\begin{abstract}

A system of diagrams is introduced that allows the representation of various
elements of a quantum circuit, including measurements, in a form which makes no
reference to time (hence ``atemporal'').  It can be used to relate quantum
dynamical properties to those of entangled states (map-state duality), and
suggests useful analogies, such as the inverse of an entangled ket.  Diagrams
clarify the role of channel kets, transition operators, dynamical operators
(matrices), and Kraus rank for noisy quantum channels.  Positive (semidefinite)
operators are represented by diagrams with a symmetry that aids in
understanding their connection with completely positive maps.  The diagrams are
used to analyze standard teleportation and dense coding, and for a careful
study of unambiguous (conclusive) teleportation. A simple diagrammatic argument
shows that a Kraus rank of 3 is impossible for a one-qubit channel modeled
using a one-qubit environment in a mixed state.

\end{abstract}


	\section{Introduction}
\label{sct1}

In some sense the problem of entangled states and the problem of noisy
channels, two central issues in quantum information theory, are one and the
same problem.  Parallels between them have been well-known for some time, at
least to those who know them.  For an extensive and extremely helpful
discussion of this duality between quantum maps and quantum states, with
copious references to earlier literature, see \cite{ZcBn04}.  The present paper
continues an effort, begun with the study of channel kets in \cite{Grff05}, to
systematize this relationship in a manner which, as far as is practical, avoids
making reference to a particular choice of basis, thus using properties of
operators rather than their (basis-dependent) matrices.  The aim is to make
these methods and the corresponding point of view more accessible to those less
familiar with their possibilities, and to classify different problems of
quantum information theory, whether solved or unsolved, in a systematic way
which reveals underlying connections.  In \cite{Grff05} a classification scheme
was proposed using properties of an entangled system at a single time, for
reasons there discussed.  The present paper, following this same motivation, is
devoted to procedures for reducing quantum circuits, together with
``preparation,'' ``measurement,'' and (at least to some extent) ``classical
communication,'' to a form in which time plays no role, using a system of what
we call \emph{atemporal diagrams}.

The atemporal form of a particular problem is not necessarily the one which
yields the best physical intuition.  Entanglement is not an easier concept than
transmission of quantum information, and we often find that a good way of
extracting physical insight from an atemporal diagram is to think of it as some
sort of quantum channel.  The diagrams do, however, provide a precise and
systematic notation for a variety of things for which Dirac notation, despite
its many advantages, is not altogether ideal; and this without introducing a
direction of time, something always present in quantum circuit diagrams.  Often
an atemporal diagram will suggest an analogy between a problem stated in terms
of how a system develops in time and another which refers to properties of a
quantum system at a single time. While some of these analogies are already
known (e.g., a noiseless quantum channel is ``like'' a fully-entangled state),
others are not so obvious.  In addition, the diagrams can speed up analysis of
some situations by allowing one to say, ``Well, that is obvious,'' without
engaging in lengthy reasoning or complicated algebra.  To be sure, the
effectiveness of diagrams of this sort is to some extent subjective.  We
ourselves have found the scheme presented in this paper to be extremely
helpful, and we hope other members of the community will share our enthusiasm
or, better yet, come up with a superior, more powerful system.

Section \ref{sct2} defines our notation for Hilbert spaces and explains the
rules for constructing diagrams, together with a small number of examples.
These include diagrams for positive (semidefinite) operators, which exhibit a
particular symmetry, ``transposers'' which change kets into maps and vice
versa, and the notion of the inverse of an entangled ket.

Applications begin in Sec.~\ref{sct3a} with a discussion of how atemporal
diagrams, and in particular the notion of the inverse of an entangled ket, can
be used to ``solve'' certain exercises in quantum circuits. The (by now)
standard protocols for teleportation and dense coding are the subject of
diagrammatic analysis in Sec.~\ref{sct3b}.

In Sec.~\ref{sct4} atemporal diagrams are applied to the problem of unambiguous
or conclusive teleportation, in which a partially-entangled pure state replaces
a fully-entangled state as the shared resource, and the agents know whether the
process succeeds or fails.  Earlier work on this problem, summarized in
Sec.~\ref{sct4e}, can be unified and generalized in a very systematic way by
using the notion of the inverse of an entangled ket. (Somewhat analogous
results on dense coding, which also make use of map-state duality but not
atemporal diagrams, will be the subject of a later publication \cite{Wuao05}.)

Noisy quantum channels are the topic of Sec.~\ref{sct5}.  In Sec.~\ref{sct5a}
diagrams corresponding to a channel ket, transition operator, and dynamical
operator are used to further illuminate the meanings of objects named in
\cite{Grff05}. (The ``dynamical operator'' corresponds to the ``dynamical
matrix'' of \cite{ZcBn04}.)  In particular, certain positivity conditions
become self-evident because of the symmetry of the diagrams, and this topic is
pursued further in Sec.~\ref{sct5b} with a discussion of completely positive
maps. With the help of diagrams one sees that the Kraus rank of a noisy quantum
channel is the same as the (ordinary) rank of a suitable ``cross operator.''
This fact is used in Sec.~\ref{sct5c} to construct a straightforward, simple
proof that the Kraus rank of a noisy one-qubit channel modeled using a
one-qubit environment in a mixed state cannot have the value 3.  Numerical
evidence for this was reported in \cite{Trao99}, but we believe ours to be the
first analytic proof.

Section~\ref{sct6} contains a summary and notes some open questions.
Appendix~\ref{sctpa}, using a mathematical result derived in App.~\ref{sctpb},
addresses some technical points about the unambiguous operations employed in
Sec.~\ref{sct4}. 

	\section{Diagrams}
\label{sct2}

	\subsection{Hilbert spaces}
\label{sct2a}

Our notation treats bras, kets, operators, entangled states, superoperators,
and the like in a uniform manner: each is an element of a suitable Hilbert
space: a complex linear space on which an inner product is defined.  Since we
will be dealing with many such spaces, part of the notational problem is making
clear precisely which space it is to which a particular entity or ``object''
belongs. We assume that all the Hilbert spaces of interest to us are
finite.

Just as in Dirac notation, where a ket $|\psi\rgl$ and a bra $\lgl\psi|$ mean
different things, we find it useful to distinguish a Hilbert space $\HS$
inhabited by kets from the dual space $\HS\ad$ inhabited by bras.  Lower-case
Roman letter subscripts are used to distinguish Hilbert spaces for different
systems, and we adopt the following abbreviations
\begin{equation}
 \HS_{ab}=\HS_a\ot\HS_b,\quad \hat\HS_a = \HS_a\ot\HS_a\ad,
 \quad \LS_{ba} = \HS_b\ot\HS_a\ad
\label{eqn1}
\end{equation}
for the \emph{tensor product} of two spaces (sums of dyads $|a\rgl\ot|b\rgl$),
the space of \emph{operators} (sums of dyads $|a\rgl\lgl a'|$), and the space
of \emph{linear maps} from $\HS_a$ to $\HS_b$, (sums of dyads $|b\rgl\lgl
a|$). Omitting the $\ot$ symbols in \eqref{eqn1} and similar expressions will
cause no harm. Note that $\HS_{ab}=\HS_{ba}$, but the order of the subscripts
in $ \LS_{ba}$ is important: it is that of the dyad $|b\rgl\lgl a|$, and also
that of the subscripts in a matrix representing such a map.  Each of these
\emph{compound} spaces, and others like them, is itself a Hilbert space, with
addition and scalar multiplication defined in the obvious way, and a scalar
inner product which for two elements $|a\rgl\ot| b\rgl\ot\lgl c|$ and $|
a'\rgl\ot| b'\rgl\ot\lgl c'|$ in $\HS_{ab}\ot\HS_c\ad$, to take a specific
example, would be $\lgl a| a'\rgl\cdot\lgl b| b'\rgl\cdot\lgl c'| c\rgl$.  The
\emph{dimension} of a Hilbert space $\HS_x$ will be denoted by $d_x$, and of
course $d_{ab}=d_a\tm d_b$.

	\subsection{Examples of diagrams}
\label{sct2b}

\begin{figure}[h]
$$
\begin{pspicture}(-0.2,-8.8)(7.9,0.5) 
\newpsobject{showgrid}{psgrid}{subgriddiv=1,griddots=10,gridlabels=6pt}
\psset{
labelsep=2.0,
arrowsize=0.150 1,linewidth=\lwd}
	\def\ahead{\mbox{
\parbox[t]{4.5cm}{\centering Diagram\\ \phantom{ddd}}%
\parbox[t]{1.5cm}{\centering Hilbert\\[-.5ex] space}
\hspace{-.2cm}
\parbox[t]{1.8cm}{\centering Dirac\\[-.5ex] notation}
	}}
	\def\aket{
\psline(0,0)(1,0)\psline{->}(0,0)(0.85,0)\rput(0,0){\obc}\rput(1,0){\opdot}
\rput(0,0){$\psi$}\rput[b](1.0,0.2){$a$}
	}
	\def\arket{%
\psline(0,0)(1,0)\psline{->}(1,0)(0.15,0)\rput(1,0){\obc}\rput(0,0){\opdot}
\rput(1,0){$\psi$}\rput[b](0,0.2){$a$}
	}
	\def\abra{
\psline(0,0)(1,0)\psline{<-}(0.55,0)(1,0)\rput(0,0){\obc}\rput(1,0){\cldot}
\rput(0,0){$\phi\ad$}\rput[b](1.0,0.2){$a$}
	}
	\def\bbra{
\psline(0,0)(1,0)\psline{<-}(0.55,0)(1,0)\rput(0,0){\obc}\rput(1,0){\cldot}
\rput(0,0){$\om\ad$}\rput[b](1.0,0.2){$b$}
	}
	\def\aopr{
\psline(0,0)(2,0)\psline{>->}(0.15,0)(1.85,0)\rput(1,0){\obc}
\rput(0,0){\cldot}\rput(2,0){\opdot}
\rput(1,0){$A$}\rput[b](0,0.2){$a$}\rput[b](2,0.2){$a$}
	}
	\def\idopr{
\psline(0,0)(2,0)\psline{>->}(0.15,0)(1.85,0)
\rput(0,0){\cldot}\rput(2,0){\opdot}
\rput[b](0,0.2){$a$}\rput[b](2,0.2){$a$}
	}
	\def\mopr{
\psline(0,0)(2,0)\psline{<-<}(0.15,0)(1.85,0)\rput(1,0){\obc}
\rput(0,0){\opdot}\rput(2,0){\cldot}
\rput(1,0){$M$}\rput[b](0,0.2){$b$}\rput[b](2,0.2){$a$}
	}
	\def\adyd{
\rput(-1.7,0){\arket}\rput(0.0,0){$\ot$}\rput(0.7,0){\abra}
	}
	\def\ketop{
\rput(-1.7,0){\mopr}\rput(0.7,0){\arket}
	}
\def\lfsb{\rput(0,0){$\left[\vrule height .4 cm depth .4 cm width 0pt\right.$}}
\def\rtsb{\rput(0,0){$\left.\vrule height .4 cm depth .4 cm width 0pt\right]$}}
	\def\bdyd{
\rput(-1.7,0){\lfsb}\rput(-1.5,0){\arket}
\rput(0.5,0){\bbra}\rput(1.7,0){\rtsb}
	}
\rput[l](0,0){\ahead}
\rput(2.5,-1){\aket} \rput[B](5.5,-1){$\HS_a$} \rput[B](7.0,-1){$|\psi\rgl$}
\rput(2.5,-2){\abra} 
\rput[B](5.5,-2){$\HS_a\ad$}\rput[B](7.0,-2){$\lgl\phi|$}
\rput(1.5,-3.0){\aopr}
\rput[B](5.5,-3){$\hat\HS_a$}\rput[B](7.0,-3){$A$}
\rput(1.5,-4){\idopr}
\rput[B](5.5,-4){$\hat\HS_a$}\rput[B](7.0,-4){$I$}
\rput(1.5,-5){\mopr}
\rput[B](5.5,-5){$\LS_{ba}$}\rput[B](7.0,-5){$M$}
\rput(2.7,-6){\adyd}
\rput[B](5.5,-6){$\hat\HS_a$}\rput[B](7.0,-6){$|\psi\rgl\lgl\phi|$}
\rput(2.7,-7.2){\bdyd}
\rput[B](5.5,-7.2){$\LS_{ab}$}\rput[B](7.0,-7.2){$|\psi\rgl\lgl\om|$}
\rput(2.5,-8.4){\ketop}
\rput[B](5.7,-8.4){$\HS_a\!\ot\!\LS_{ba}$}
\rput[B](7.3,-8.4){$M\!\ot\!|\psi\rgl$}
\rput[l](0,-1){(a)}
\rput[l](0,-2){(b)}
\rput[l](0,-3){(c)}
\rput[l](0,-4){(d)}
\rput[l](0,-5){(e)}
\rput[l](0,-6){(f)}
\rput[l](0,-7.2){(g)}
\rput[l](0,-8.4){(h)}
\end{pspicture}
$$
\caption{Examples of simple diagrams (a-e) and product diagrams (f-h).}
\label{fgr1}
\end{figure}

Each ``object'' which is an element of some Hilbert space can be denoted by a
diagram in which a symbol representing it is placed inside a circle or square
or some other shape forming the \emph{center}, connected by a certain number of
\emph{legs}, straight or curved lines, to \emph{nodes} which specify the
relevant Hilbert space.  Several examples are shown Fig.~\ref{fgr1}.  A node is
\emph{open} $\circ$ or \emph{closed} $\bullet$ depending on whether it refers
to the Hilbert space or its dual, and is identified by a letter placed near it.
Sometimes we shall refer to these as \emph{active} nodes, in contrast to 
inactive nodes present on internal lines of a diagram following contraction;
the latter are generally not shown on the diagram (see Sec.~\ref{sct2c}).
In addition to the nodes, there are arrows on the legs pointing outward from
the center towards open nodes, and inward from closed nodes towards the center.
The identity operator $I$ is indicated by a single line, see (d), without a
center; this notation will be justified in Sec.~\ref{sct2c}.

The Dirac notation, shown in the third column of Fig.~\ref{fgr1}, resembles
that in the diagrams, with an important exception.  The bra symbol $\lgl\phi|$
corresponds to an object with the label $\phi\ad$ rather than $\phi$; were the
object labeled $\phi$, the bra symbol would instead be $\lgl\phi\ad|$. One can
think of the Dirac $\lgl\,|$ as effectively placing a dagger on the symbol it
contains.

There is no rule that prescribes the orientations of the legs emerging from the
center of an object; these may be chosen as convenient, as illustrated in the
opposite orientations shown in parts (c) and (e) of the figure.  Of course, if
there are several legs, restricting oneself to rigid rotations will make visual
identification simpler if the object is used more than once (e.g., the box
labeled $V$ in Fig.~\ref{fgr11} below).  But this is not an absolute rule, as
the symbol in the center and the labels on the nodes or legs remove any
ambiguity.

Objects which are tensor products of entities on distinct Hilbert spaces can be
drawn as disconnected pieces, illustrated in parts (f) to (h) of
Fig.~\ref{fgr1}. Sometimes it may be helpful to emphasize that different
disconnected pieces belong to the same diagram by placing a $\ot$ between
adjacent pieces (f), or enclosing the pieces inside a large pair of
enclosing brackets (g).  However, there is no rule that these
constructions have to be used, and they can be omitted when ambiguity is
unlikely, as in (h).  There is also no rule that says that tensor products
must be represented as disconnected pieces; the one in (h), for example,
could be represented using a single center and three legs terminating in open
$a$ and $b$ nodes and a closed $a$ node.

There is some redundancy in using both arrows and nodes, and one or the other
could be omitted without changing the significance of a diagram.  Retaining the
nodes makes it easier to distinguish legs from other things in a complicated
diagram.  On the other hand, it is convenient to eliminate them when carrying
out contractions, see below.

	\subsection{Contractions and matrices}
\label{sct2c}

\begin{figure}[h]
$$
\begin{pspicture}(0,-9.5)(8,0.6) 
\newpsobject{showgrid}{psgrid}{subgriddiv=1,griddots=10,gridlabels=6pt}
\psset{
labelsep=2.0,
arrowsize=0.150 1,linewidth=\lwd}
\def\dput(#1)#2#3{\rput(#1){#2}\rput(#1){#3}} 
\def\clnr(#1)#2{\dput(#1){\cldot}{\rput[r](-0.2,0){$#2$}}} 
\def\clnl(#1)#2{\dput(#1){\cldot}{\rput[l](0.2,0){$#2$}}} 
\def\clnb(#1)#2{\dput(#1){\cldot}{\rput[b](0,0.2){$#2$}}} 
\def\clnt(#1)#2{\dput(#1){\cldot}{\rput[t](0,-0.2){$#2$}}} 
\def\opnr(#1)#2{\dput(#1){\opdot}{\rput[r](-0.2,0){$#2$}}} 
\def\opnl(#1)#2{\dput(#1){\opdot}{\rput[l](0.2,0){$#2$}}} 
\def\opnb(#1)#2{\dput(#1){\opdot}{\rput[b](0,0.2){$#2$}}} 
\def\opnt(#1)#2{\dput(#1){\opdot}{\rput[t](0,-0.2){$#2$}}} 
\def\rarr(#1){\rput(#1){\psline{->}(-0.2,0)(0,0)}}
\def\larr(#1){\rput(#1){\psline{->}(0.2,0)(0,0)}}
\def\uarr(#1){\rput(#1){\psline{->}(0,-0.2)(0,0)}}
\def\darr(#1){\rput(#1){\psline{->}(0,0.2)(0,0)}}
	\def\hedder{\mbox{
\hspace{0.3cm}
\parbox[t]{2cm}{\centering Initial\\[-.5ex] object}%
\parbox[t]{3.7cm}{\centering After contraction\\[-.5ex] on $\HS_a$}
\hspace{-.4cm}
\parbox[t]{1.8cm}{\centering Dirac\\[-.5ex] notation}
	}}
	\def\aket{
\psline(0,0)(1,0)\psline{->}(0,0)(0.85,0)\rput(0,0){\obc}\rput(1,0){\opdot}
\rput(0,0){$\psi$}\rput[b](1.0,0.2){$a$}
	}
	\def\abra{
\psline(0,0)(1,0)\psline{<-}(0.55,0)(1,0)\rput(0,0){\obc}\rput(1,0){\cldot}
\rput(0,0){$\phi\ad$}\rput[t](1.0,-0.2){$a$}
	}
	\def\acase{
\rput(0,0){\aket} \rput(0,-1){\abra}\rput(0.5,-0.5){$\ot$} 
	}
	\def\acontra{
\rput(0,0){\aket} \rput(0,-1){\abra}
\psline(1,0)(1,-1)\psline{->}(1,0)(1,-0.7)
	}
	\def\acontrb{
\psarc(-0.3,-0.5){.8}{-60}{60}\psarc{-<}(-0.3,-0.5){.8}{-60}{0}
\rput(0,0){\obc}\rput(0,0){$\psi$}
\rput(0,-1){\obc}\rput(0,-1){$\phi\ad$}
\rput[l](0.65,-.5){$a$}
	}
	\def\bcontra{
\psline(0,0)(2,0)\psline{>->}(0.15,0)(1.85,0)
\psline(2,0)(2,-.6)(0,-.6)(0,0)\psline{<-}(0.85,-.6)(2,-.6)
\rput(1,0){\obc}\rput(0,0){\cldot}\rput(2,0){\opdot}
\rput(1,0){$A$}\rput[b](0,0.2){$a$}\rput[b](2,0.2){$a$}
	}
	\def\oprv{
\psline(0,0)(0,2)\psline{>->}(0,0.15)(0,1.85)
\rput(0,0){\cldot}\rput(0,2){\opdot}
\rput(0,1){\obc}\rput(0,1){$A$}
\rput[r](-.2,0){$a$}\rput[r](-.2,2){$a$}
	}
	\def\bcontra{
\psline(0,0)(0,2)\psline{>->}(0,0.15)(0,1.85)
\psline(0,0)(0.6,0)\psline(0,2)(0.6,2)\psline(0.6,0)(0.6,2)
\psline{-<}(0.6,0)(0.6,1.1)
\rput(0,0){\cldot}\rput(0,2){\opdot}
\rput(0,1){\obc}\rput(0,1){$A$}
\rput[r](-.2,0){$a$}\rput[r](-.2,2){$a$}
	}
	\def\bcontrb{
\psbezier{->}(0,1)(0,2.2)(0.7,2.2)(0.7,1)
\psbezier(0,1)(0,-.2)(0.7,-.2)(0.7,1.1)
\rput(0,1){\obc}\rput(0,1){$A$}
\rput[l](.8,1){$a$}
	}
	\def\cobja{
\psline(0,0)(1.6,0)
\dput(0.8,0){\obc}{$M$}
\larr(0.15,0)\larr(1.2,0)
\opnb(0,0){b}
\clnb(1.6,0){a}
\rput(0.3,-1.1){\aket} \rput(0.55,-.55){$\ot$}
	}
	\def\cobjb{
\psline(0,0)(2.0,0)
\dput(0.8,0){\obc}{$M$}
\dput(2.0,0){\obc}{$\psi$}
\opnb(0,0){b}
\rput[b](1.4,0.2){$a$}
\larr(0.15,0)\larr(1.3,0)
	}
	\def\cobjc{
\psline(0,0)(0.8,0)
\dput(1.0,0){\obr{0.5}}{$M\psi$}
\opnb(0,0){b}
\larr(0.15,0)
	}
	\def\dobja{
\psline(0,0.5)(2.2,0.5)
\psline(0,-0.5)(2.2,-0.5)
\dput(1.1,0){\bobx}{$W$}
\opnb(0,0.5){a}\clnt(0,-0.5){a}
\opnb(2.2,0.5){b}\clnt(2.2,-0.5){b}
\larr(0.2,0.5)\rarr(0.4,-0.5)
\rarr(2.0,0.5)\larr(1.8,-0.5)
	}
	\def\dobjb{
\psline(1,0.5)(2.2,0.5)
\psline(1,-0.5)(2.2,-0.5)
\psbezier{->}(0.6,0.5)(0.1,0.5)(0,0.4)(0,-0.2)
\psbezier(0.6,-0.5)(0.1,-0.5)(0,-0.4)(0,0.1)
\rput[r](-0.15,0){$a$}
\dput(1.1,0){\bobx}{$W$}
\opnb(2.2,0.5){b}\clnt(2.2,-0.5){b}
\rarr(2.0,0.5)\larr(1.8,-0.5)
	}
\rput[l](0,0.2){\hedder}

\rput[l](0,-1.5){(a)}
\rput(1.1,-1){\acase}
\rput(3.2,-1){\acontra}
\rput(4.7,-1.5){$=$}
\rput(5.2,-1){\acontrb}
\rput(7.0,-1.5){$\lgl\phi|\psi\rgl$}

\rput[l](0,-4){(b)}
\rput(1.3,-5){\oprv}
\rput(3,-5){\bcontra}
\rput(4.8,-5){\bcontrb}
\rput(7,-4){$\Tr(A)$}
\rput(4.1,-4){$=$}

\rput[l](0,-6.5){(c)}
\rput(0.8,-6){\cobja}
\rput(3.3,-6){\cobjb}
\rput(3.4,-7){$=$}
\rput(3.8,-7){\cobjc}
\rput(7,-6.5){$M|\psi\rgl$}

\rput[l](0,-8.5){(d)}
\rput(0.7,-8.5){\dobja}
\rput(3.7,-8.5){\dobjb}
\rput(7,-8.5){$\Tr_a(W)$}
\end{pspicture}
$$
\caption{ Examples illustrating contraction}
\label{fgr2}
\end{figure}
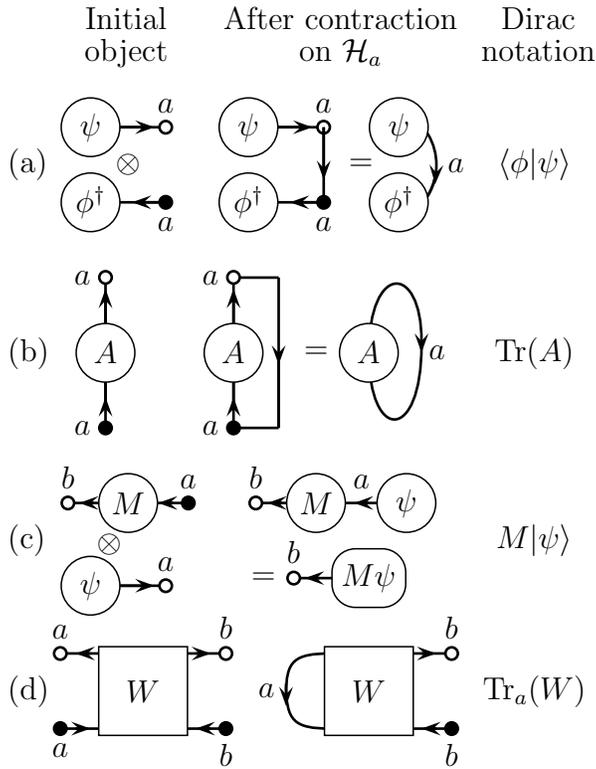

A \emph{contraction}, meaning an inner product, trace, or partial trace, can be
carried out on any object whose diagram contains both a closed and an open node
referring to the same Hilbert space, i.e., to the Hilbert space and its
dual. It is performed by drawing a line connecting these two nodes, with an
arrow pointing from the open to the closed node.  There are several examples in
Fig.~\ref{fgr2}.  As the contracted object makes no reference to the Hilbert
space of the contracted nodes, they can be omitted, and the three arrows
pointing in the same direction replaced by just one.  It is useful to place a
single letter next to the remaining arrow on the contracting line to indicate
on which space the contraction has been carried out.  (The original nodes can
be put back in, if desired, but they should then not be confused with the
\emph{active} nodes, which always terminate a single line.) This process of
simplification is indicated explicitly in Fig.~\ref{fgr2}(a) and (b), but only
the end result in (c) and (d).  (One could also leave the two nodes joined by
the contracting line and omit the arrows, but this is a bulkier notation.)
Using the abbreviation just mentioned justifies the use of a single line to
indicate the identity operator, Fig.~\ref{fgr1} (d), as can be seen by
constructing the diagram for $I|\psi\rgl=|\psi\rgl$.  Of course, both arrows
and labels are sometimes omitted when the context determines what they should
be.  Figure~\ref{fgr2}(c) illustrates still another abbreviation which is
often, but not always, convenient.  Two centers connected by a single
contraction line can be reduced to a single center containing the symbol
corresponding to the head of the arrow to the left of that corresponding to the
tail.  The order of these symbols is important, and (usually) agrees with
standard notation.

\begin{figure}[h]
$$
\begin{pspicture}(0,-11.3)(8,0.7) 
\newpsobject{showgrid}{psgrid}{subgriddiv=1,griddots=10,gridlabels=6pt}
\psset{
labelsep=2.0,
arrowsize=0.150 1,linewidth=\lwd}
\def\dput(#1)#2#3{\rput(#1){#2}\rput(#1){#3}} 
\def\obcs{\pscircle[fillcolor=white,fillstyle=solid,linewidth=\lwn](0,0){0.35}}
\def\clnr(#1)#2{\dput(#1){\cldot}{\rput[r](-0.2,0){$#2$}}} 
\def\clnl(#1)#2{\dput(#1){\cldot}{\rput[l](0.2,0){$#2$}}} 
\def\clnb(#1)#2{\dput(#1){\cldot}{\rput[b](0,0.2){$#2$}}} 
\def\clnt(#1)#2{\dput(#1){\cldot}{\rput[t](0,-0.2){$#2$}}} 
\def\opnr(#1)#2{\dput(#1){\opdot}{\rput[r](-0.2,0){$#2$}}} 
\def\opnl(#1)#2{\dput(#1){\opdot}{\rput[l](0.2,0){$#2$}}} 
\def\opnb(#1)#2{\dput(#1){\opdot}{\rput[b](0,0.2){$#2$}}} 
\def\opnt(#1)#2{\dput(#1){\opdot}{\rput[t](0,-0.2){$#2$}}} 
\def\rarr(#1){\rput(#1){\psline{->}(-0.2,0)(0,0)}}
\def\larr(#1){\rput(#1){\psline{->}(0.2,0)(0,0)}}
\def\uarr(#1){\rput(#1){\psline{->}(0,-0.2)(0,0)}}
\def\darr(#1){\rput(#1){\psline{->}(0,0.2)(0,0)}}

	\def\hedder{\mbox{
\parbox[t]{2cm}{\centering Object}%
\hspace{0.3cm}
\parbox[t]{3.7cm}{\centering Matrix}
\parbox[t]{1.4cm}{\centering Dirac\\[-.5ex] notation}
	}}
	\def\aobja{
\psline(0,0)(1,0)\psline{->}(0,0)(0.85,0)\rput(0,0){\obc}\rput(1,0){\opdot}
\rput(0,0){$\psi$}\rput[b](1.0,0.2){$a$}
	}
	\def\aobjb{
\psline(0,0)(1.5,0)\psline{-<}(0,0)(0.8,0)
\rput(0,0){\obc}\rput(0,0){$a_j\ad$}
\rput(1.5,0){\obc}\rput(1.5,0){$\psi$}
	}
	\def\aobjc{
\rput(0,0){\obr{0.5}}\rput(0,0){$a_j\ad\psi$}
	}
	\def\bobja{
\psline(0,0)(0,2)\psline{>->}(0,0.15)(0,1.85)
\rput(0,0){\cldot}\rput(0,2){\opdot}
\rput(0,1){\obc}\rput(0,1){$M$}
\rput[r](-.2,0){$a$}\rput[r](-.2,2){$b$}
	}
	\def\bobjb{
\psline(0,0)(2.4,0)\psline{->}(0,0)(0.7,0)\psline{->}(1.0,0)(1.9,0)
\rput(0,0){\obc}\rput(0,0){$a_j$}
\rput(1.2,0){\obc}\rput(1.2,0){$M$}
\rput(2.4,0){\obc}\rput(2.4,0){$b_k\ad$}
	}
	\def\cobja{
\psline(0,1)(0,0)\psline{-<}(0,1)(0,0.15)
\psline(0,1)(-.867,1.5)\psline{->}(0,1)(-.737,1.425)
\psline(0,1)(.867,1.5)\psline{->}(0,1)(.737,1.425)
\rput(0,1){\obc}\rput(0,1){$Y$}
\clnl(0,0){c}
\opnb(-.867,1.5){a}
\opnb(.867,1.5){b}
	}
	\def\cobjb{
\psline(0,1)(0,0)\psline{-<}(0,1)(0,0.15)
\psline(0,1)(-1.212,1.7)\psline{->}(0,1)(-.737,1.425)
\psline(0,1)(.867,1.5)\psline{->}(0,1)(.737,1.425)
\rput(0,1){\obc}\rput(0,1){$Y$}
\dput(-1.212,1.7){\obc}{$a_j\ad$}
\rput[b](-0.5,1.6){$a$}
\opnb(.867,1.5){b}
\clnl(0,0){c}
	}
	\def\dobjb{
\psline(0,1)(0,-.5)
\uarr(0,0.5)
\psline(0,1)(-1.212,1.7)\psline{->}(0,1)(-.737,1.425)
\psline(0,1)(.867,1.5)\psline{->}(0,1)(.737,1.425)
\rput(0,1){\obc}\rput(0,1){$Y$}
\dput(0,-.2){\obc}{$c_l$}
\dput(-1.212,1.7){\obc}{$a_j\ad$}
\rput[l](0.2,0.4){$c$}
\rput[b](-0.5,1.6){$a$}
\opnb(.867,1.5){b}
	}
	\def\eobja{
\psline(0,0)(0,1)
\clnr(0,0){a}
\opnr(0,1){a}
\uarr(0,0.65)
	}
	\def\eobjb{
\psline(0,0)(1.2,0)
\psline(0,1)(1.2,1)
\dput(1.2,0){\obc}{$a_j\ad$}
\dput(1.2,1){\obc}{$a_j$}
\opnr(0,1){a}
\clnr(0,0){a}
\rarr(0.55,0)
\larr(0.3,1)
\rput(0.7,0.5){$\ot$}
	}

\rput[l](0,0.2){\hedder}

\rput[l](0,-1){(a)}
	\rput(0.4,0){
\rput(0.7,-1){\aobja}
\rput(2.6,-1){\aobjb}
\rput(4.8,-1){$=$}
\rput(5.6,-1){\aobjc}
\rput(7,-1){$\lgl a_j | \psi\rgl$}
	}
\rput[l](0,-3){(b)}
	\rput(0.2,0){
\rput(1,-4){\bobja}
\rput(2.6,-3){\bobjb}
\rput(6.9,-3){$\lgl b_k|M|a_j\rgl$}
	}
\rput[l](0,-5.5){(c)}
\rput(1.2,-6.5){\cobja}
\rput(4.5,-6.5){\cobjb}
\rput(7,-5.5){$\lgl a_j|Y$}

\rput[l](0,-8){(d)}
\rput(1.2,-9){\cobja}
\rput(4.5,-8.7){\dobjb}
\rput(7,-8){$\lgl a_j|Y|c_l \rgl$}

\rput[l](0,-10.5){(e)}
\rput(1,-11){\eobja}
\rput[l](1.3,-10.5){$=\sum_j$}
\rput(2.9,-11){\eobjb}
\rput(6.5,-10.5){$I=\sum_j |a_j\rgl\lgl a_j|$}

\end{pspicture}
$$
\caption{Parts (a) through (d) are examples of matrix elements of different
  objects, together with the corresponding Dirac notation; (e) is the
  diagrammatic form of $I$ as a sum of dyads.}
\label{fgr3}
\end{figure}
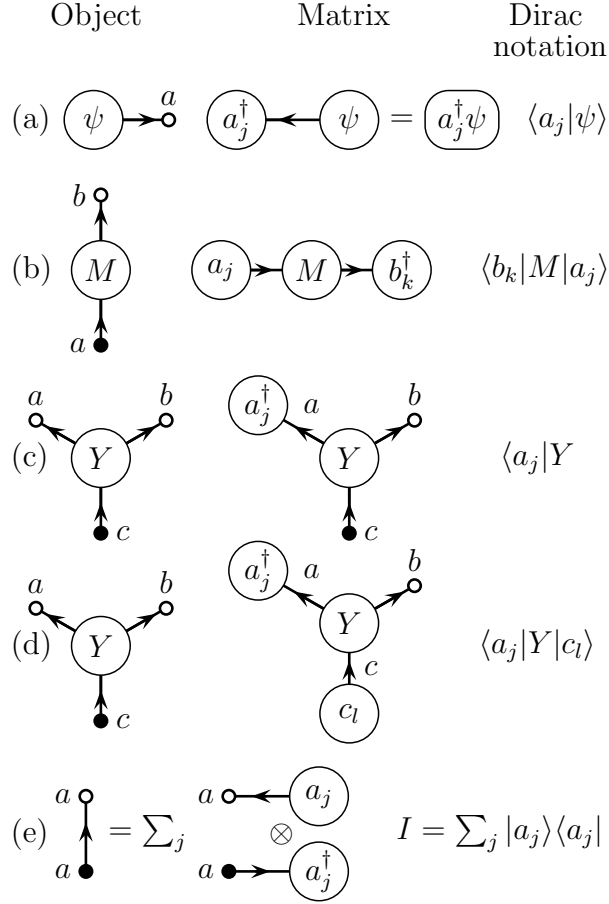

To obtain the \emph{matrix} associated with some object, it is necessary to
specify a basis or bases.  We limit ourselves to orthonormal bases, although
the same diagrammatic approach will work for more general choices.  Examples
are shown in Fig.~\ref{fgr3}, together with the corresponding Dirac notation
for the matrix. Note that column vectors, row vectors, and objects labeled by
three or more indices are included in the scheme, and the elements of a
rectangular matrix may turn out to be operators or matrices, as in (d).  The
Dirac notation for items (c) and (d), while appropriate, can sometimes mislead;
e.g., $\lgl a_j|Y|c_l\rgl$ denotes a ket, not a complex number.

	\subsection{Adjoints and positive operators}
\label{sct2d}

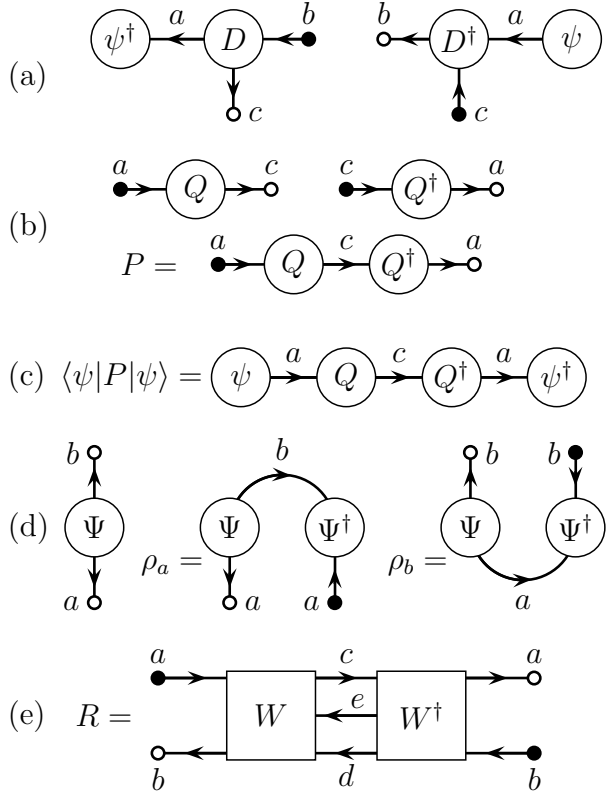
\begin{figure}[h]
$$
\begin{pspicture}(0,-10.0)(8.0,0.6) 
\newpsobject{showgrid}{psgrid}{subgriddiv=1,griddots=10,gridlabels=6pt}
\psset{
labelsep=2.0,
arrowsize=0.150 1,linewidth=\lwd}
\def\dput(#1)#2#3{\rput(#1){#2}\rput(#1){#3}} 
\def\obcs{\pscircle[fillcolor=white,fillstyle=solid,linewidth=\lwn](0,0){0.35}}
\def\clnr(#1)#2{\dput(#1){\cldot}{\rput[r](-0.2,0){$#2$}}} 
\def\clnl(#1)#2{\dput(#1){\cldot}{\rput[l](0.2,0){$#2$}}} 
\def\clnb(#1)#2{\dput(#1){\cldot}{\rput[b](0,0.2){$#2$}}} 
\def\clnt(#1)#2{\dput(#1){\cldot}{\rput[t](0,-0.2){$#2$}}} 
\def\opnr(#1)#2{\dput(#1){\opdot}{\rput[r](-0.2,0){$#2$}}} 
\def\opnl(#1)#2{\dput(#1){\opdot}{\rput[l](0.2,0){$#2$}}} 
\def\opnb(#1)#2{\dput(#1){\opdot}{\rput[b](0,0.2){$#2$}}} 
\def\opnt(#1)#2{\dput(#1){\opdot}{\rput[t](0,-0.2){$#2$}}} 
\def\rarr(#1){\rput(#1){\psline{->}(-0.2,0)(0,0)}}
\def\larr(#1){\rput(#1){\psline{->}(0.2,0)(0,0)}}
\def\uarr(#1){\rput(#1){\psline{->}(0,-0.2)(0,0)}}
\def\darr(#1){\rput(#1){\psline{->}(0,0.2)(0,0)}}
	\def\aobj{
\psline(0,0)(2.5,0)\psline{<-}(0.6,0)(1.5,0)\psline{<-}(2.05,0)(2.5,0)
\rput[b](0.75,0.2){$a$}
\psline(1.5,-1)(1.5,0)\psline{<-}(1.5,-0.8)(1.5,0)
\rput(0,0){\obc}\rput(0,0){$\psi\ad$}
\rput(1.5,0){\obc}\rput(1.5,0){$D$}
\rput(2.5,0){\cldot}\rput[b](2.5,0.2){$b$}
\rput(1.5,-1){\opdot}\rput[l](1.7,-1){$c$}
	}
	\def\aobjad{
\psline(0,0)(2.5,0)\psline{<-}(0.2,0)(1,0)\psline{<-}(1.6,0)(2.5,0)
\rput[b](1.75,0.2){$a$}
\psline(1,-1)(1,0)\psline{<-}(1,-0.55)(1,-.9)
\rput(1,0){\obc}\rput(1,0){$D\ad$}
\rput(2.5,0){\obc}\rput(2.5,0){$\psi$}
\rput(0,0){\opdot}\rput[b](0,0.2){$b$}
\rput(1,-1){\cldot}\rput[l](1.2,-1){$c$}
	}
	\def\bobja{
\psline(0,0)(2,0)\psline{->}(0,0)(0.45,0)\psline{->}(1.0,0)(1.8,0)
\rput(0,0){\cldot}\rput[b](0,0.2){$a$}
\rput(2,0){\opdot}\rput[b](2,0.2){$c$}
\rput(1,0){\obc}\rput(1,0){$Q$}
	}
	\def\bobjb{
\psline(0,0)(2,0)\psline{->}(0,0)(0.45,0)\psline{->}(1.0,0)(1.8,0)
\rput(0,0){\cldot}\rput[b](2,0.2){$a$}
\rput(2,0){\opdot}\rput[b](0,0.2){$c$}
\rput(1,0){\obc}\rput(1,0){$Q\ad$}
	}
	\def\bobjc{
\psline(0,0)(3.4,0)
\psline{->}(0,0)(0.45,0)\psline{->}(3,0)(3.2,0)
\psline{->}(1,0)(1.85,0)\rput[b](1.7,0.2){$c$}
\rput(1,0){\obc}\rput(1,0){$Q$}
\rput(2.4,0){\obc}\rput(2.4,0){$Q\ad$}
\rput(0,0){\cldot}\rput[b](0,0.2){$a$}
\rput(3.4,0){\opdot}\rput[b](3.4,0.2){$a$}
	}
	\def\cobj{
\psline(0,0)(4.2,0)
\psline{->}(0,0)(0.85,0)\rput[b](0.7,0.2){$a$}
\psline{->}(0,0)(2.25,0)\rput[b](2.1,0.2){$c$}
\psline{->}(0,0)(3.65,0)\rput[b](3.5,0.2){$a$}
\rput(0,0){\obc}\rput(0,0){$\psi$}
\rput(1.4,0){\obc}\rput(1.4,0){$Q$}
\rput(2.8,0){\obc}\rput(2.8,0){$Q\ad$}
\rput(4.2,0){\obc}\rput(4.2,0){$\psi\ad$}
	}
	\def\dobja{
\psline(0,0)(0,2)
\dput(0,1){\obc}{$\Psi$}
\darr(0,0.2)
\uarr(0,1.8)
\opnr(0,0){a}
\opnr(0,2){b}
	}
	\def\dobjb{
\psline(0,0)(0,1)
\darr(0,0.2)
\psline(1.4,0)(1.4,1)
\uarr(1.4,0.45)
\psarc(0.7,1){0.7}{0}{180}
\psarc{<-}(0.7,1){0.7}{80}{180}
\dput(0,1){\obc}{$\Psi$}
\dput(1.4,1){\obc}{$\Psi\ad$}
\opnl(0,0){a}
\clnr(1.4,0){a}
\rput[b](0.7,1.9){$b$}
	}
	\def\dobjc{
\psline(0,2)(0,1)
\uarr(0,1.8)
\psline(1.4,2)(1.4,1)
\darr(1.4,1.55)
\psarc(0.7,1){0.7}{180}{0}
\psarc{->}(0.7,1){0.7}{180}{280}
\dput(0,1){\obc}{$\Psi$}
\dput(1.4,1){\obc}{$\Psi\ad$}
\opnl(0,2){b}
\clnr(1.4,2){b}
\rput[t](0.7,0.1){$a$}
	}
	\def\eobj{
\psline(0,0.5)(5.0,0.5)
\rput(0,0.5){\cldot}\rput[b](0,0.7){$a$}
\rput(5,0.5){\opdot}\rput[b](5,0.7){$a$}
\psline{->}(0,0.5)(0.65,0.5)\psline{->}(4,0.5)(4.65,0.5)
\psline{->}(1,0.5)(2.65,0.5)\rput[b](2.5,0.7){$c$}
\psline(1.5,0)(3.5,0)\psline{<-}(2.2,0)(3.5,0)\rput[b](2.65,0.1){$e$}
\psline(0,-0.5)(5.0,-0.5)
\rput(0,-0.5){\opdot}\rput[t](0,-0.7){$b$}
\rput(5,-0.5){\cldot}\rput[t](5,-0.7){$b$}
\psline{<-}(0.35,-0.5)(1.5,-0.5)\psline{<-}(4.35,-0.5)(5,-0.5)
\psline{<-}(2.35,-0.5)(3,-0.5)\rput[t](2.5,-0.65){$d$}
\rput(1.5,0){\bobx}\rput(1.5,0){$W$}
\rput(3.5,0){\bobx}\rput(3.5,0){$W\ad$}
	}
\rput[l](0,-.5){(a)}
\rput(1.5,0){\aobj}
\rput(5.0,0){\aobjad}
\rput[l](0,-2.5){(b)}
\rput(1.5,-2){\bobja}
\rput(4.5,-2){\bobjb}
\rput[l](1.5,-3){$P=$}
\rput(2.8,-3){\bobjc}
\rput[l](0,-4.5){(c)}
\rput[l](.7,-4.5){$\lgl\psi|P|\psi\rgl=$}
\rput(3.1,-4.5){\cobj}

	\rput(0,-9){
\rput[l](0,0){(e)}
\rput(2,0){\eobj}
\rput[l](.9,0){$R=$}
	}
	\rput(0.15,-7.5){
\rput[l](-0.15,1){(d)}
\rput(1,0){\dobja}
\rput[r](2.45,0.5){$\rho_a=$}
\rput(2.8,0){\dobjb}
\rput[r](5.7,0.5){$\rho_b=$}
\rput(6,0){\dobjc}
	}

\end{pspicture}
$$
\caption{ (a) The two objects are adjoints of each other. (b) A positive
operator can be formed by contraction of an object with its adjoint.  (c)
Diagram symmetry illustrates positivity criterion for $P$. (d) Reduced density
operators for a bipartite ket. (e) Positivity of $R$ follows from symmetry of
diagram. }
\label{fgr4}
\end{figure}

Given any object $O$ on a tensor product of Hilbert spaces, its adjoint $O\ad$
is defined in the usual way using an antilinear map in which all bras become
kets and vice versa.  For example, the dagger in
\begin{equation}
\Bigl[\sum c_{jkl} \Blp |a_j\rgl\ot\lgl a_k|\ot|b_l\rgl\Brp\Bigr]\ad
 = \sum c^*_{jkl} \Blp\lgl a_j|\ot | a_k\rgl\ot\lgl b_l|\Brp
\label{eqn2}
\end{equation}
maps an object from $\hat\HS_a\HS_b$ to $\hat\HS_a\HS_b\ad$.  Normally when
using Dirac notation one would omit the first $\ot$ on both the left and right
sides of this equation, by using $|a_j\rgl\lgl a_k|$ and $|a_k\rgl\lgl a_j|$,
respectively.  The rule for diagrams is illustrated in Fig.~\ref{fgr4}(a),
where the right and left sides are adjoints of each other: change every open
node to a closed node and vice versa while keeping the node labels the same,
reverse the direction of every arrow, and place the adjoint superscript
$\dagger$ on the symbol appearing in each center, unless there is one already
there, in which case remove it.  Note the usual rules, that when $\dagger$ is
applied twice to any symbol one gets the symbol back again, and if a symbol is
a product, as $M\psi$ in the abbreviated form in Fig.~\ref{fgr2}(c), one
should reverse the order: $(M\psi)\ad = \psi\ad M\ad$.  In Fig.~\ref{fgr4}(a),
the adjoint diagram has also been reflected about a vertical axis.  While this
is often a convenient thing to do, especially when one intends to carry out a
contraction to produce a positive operator, see below, it is \emph{not}
essential: the significance of any diagram does not depend upon its orientation
or the order of the legs (assuming they are properly labeled).

By a \emph{positive operator} on a Hilbert space $\HS$ we mean a Hermitian
operator in $\hat\HS$ whose eigenvalues are all nonnegative.  The longer term
``positive semidefinite'' is more precise.  (It is important to distinguish a
positive operator in the sense just defined from a \emph{positive} or
\emph{completely positive superoperator}, referring to a map from a space of
operators, say $\hat\HS_a$ to a different (or possibly the same) space of
operators $\hat\HS_b$; see Sec.~\ref{sct5}.)  A positive operator $P$
has a positive square root, and hence it can be always be written in the form
\begin{equation}
 P=Q\ad Q
\label{eqn3}
\end{equation}
by setting $Q=Q\ad$ equal to this square root.  Conversely, any operator $P$ of
the form \eqref{eqn3} is necessarily positive, even in cases in which $Q$ is a
map of $\HS$ to another Hilbert space $\HS'$, in which case $Q\ad$ maps from
$\HS'$ back to $\HS$.

The diagram representing the right side of \eqref{eqn3} is formed by a
contraction of the diagram for $Q$ with that of its adjoint,
Fig.~\ref{fgr4}(b), and exhibits a characteristic symmetry which is often
useful for visually identifying positive operators: reflection across a
particular plane changes closed to open nodes and vice versa, reverses the
directions of all arrows, and applies a superscript $\dagger$ to all symbols.
Part (c) of the figure shows how the well-known characterization of a positive
operator by the condition that $\lgl\psi|P|\psi\rgl$ be positive finds a
natural expression in diagrammatic terms when $P$ is of the form \eqref{eqn3}.
Diagrams corresponding to the reduced density operators
\begin{equation}
  \rho_a = \Tr_b\Blp|\Psi\rgl\lgl\Psi|\Brp,\quad
  \rho_b = \Tr_a\Blp|\Psi\rgl\lgl\Psi|\Brp
\label{eqn4}
\end{equation}
of an entangled ket $|\Psi\rgl\in\HS_{ab}$ are shown in (d). Finally, a more
complicated example is given in (e), where the structure of the diagram shows
that $R$ is a positive operator on $\LS_{ba}$, i.e., as a map of the linear
space of $\HS_a$-to-$\HS_b$ maps onto itself.  (Think of $R$ as ``operating''
through contraction of the left pair of nodes with something in $\LS_{ba}$ in
order to produce something else in this space, corresponding to the right pair
of nodes.)  The symmetry which shows it to be positive is reflection about a
line bisecting the $c$, $d$, and $e$ lines; note that not all of the arrows
need go in the same direction. (One may also think of $R$ as a
``top-to-bottom'' map of $\hat\HS_a$ to $\hat\HS_b$. As such it is not a
positive operator, since the domain and range are different, though it is, as
will be discussed in Sec.~\ref{sct5}, a completely positive superoperator.)

	\subsection{Transpose, rank, inverse}
\label{sct2e}

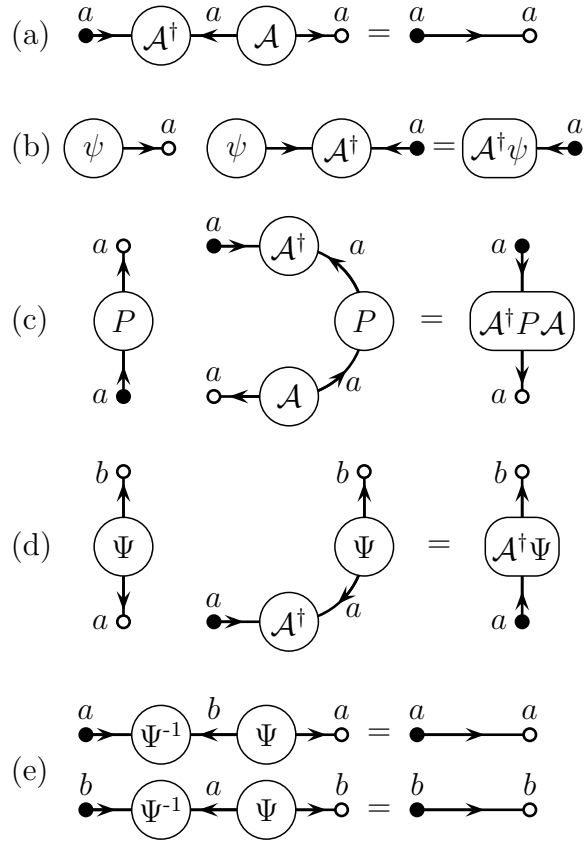
\begin{figure}[h]
$$
\begin{pspicture}(0,-10.3)(8.0,0.8) 
\newpsobject{showgrid}{psgrid}{subgriddiv=1,griddots=10,gridlabels=6pt}
\psset{
labelsep=2.0,
arrowsize=0.150 1,linewidth=\lwd}
\def\mnone{{\hbox{\scriptsize-1}}}
\def\mnonead{{\hbox{\scriptsize-1$\dagger$}}}
\def\dput(#1)#2#3{\rput(#1){#2}\rput(#1){#3}} 
\def\obcs{\pscircle[fillcolor=white,fillstyle=solid,linewidth=\lwn](0,0){0.35}}
\def\lfsb{\rput(0,0){$\left[\vrule height 1.0cm depth 1.0cm width 0pt\right.$}}
\def\rtsb{\rput(0,0){$\left.\vrule height 1.0cm depth 1.0cm width 0pt\right]$}}
\def\clnr(#1)#2{\dput(#1){\cldot}{\rput[r](-0.2,0){$#2$}}} 
\def\clnl(#1)#2{\dput(#1){\cldot}{\rput[l](0.2,0){$#2$}}} 
\def\clnb(#1)#2{\dput(#1){\cldot}{\rput[b](0,0.2){$#2$}}} 
\def\clnt(#1)#2{\dput(#1){\cldot}{\rput[t](0,-0.2){$#2$}}} 
\def\opnr(#1)#2{\dput(#1){\opdot}{\rput[r](-0.2,0){$#2$}}} 
\def\opnl(#1)#2{\dput(#1){\opdot}{\rput[l](0.2,0){$#2$}}} 
\def\opnb(#1)#2{\dput(#1){\opdot}{\rput[b](0,0.2){$#2$}}} 
\def\opnt(#1)#2{\dput(#1){\opdot}{\rput[t](0,-0.2){$#2$}}} 
\def\rarr(#1){\rput(#1){\psline{->}(-0.2,0)(0,0)}}
\def\larr(#1){\rput(#1){\psline{->}(0.2,0)(0,0)}}
\def\uarr(#1){\rput(#1){\psline{->}(0,-0.2)(0,0)}}
\def\darr(#1){\rput(#1){\psline{->}(0,0.2)(0,0)}}
	\def\aobja{
\psline(0,0)(3.4,0)
\psline{->}(0,0)(0.4,0)\psline{->}(0,0)(3.2,0)
\psline{<-}(1.5,0)(2,0)
\rput[b](1.7,0.2){$a$}
\rput(0,0){\cldot}\rput[b](0,0.2){$a$}
\rput(3.4,0){\opdot}\rput[b](3.4,0.2){$a$}
\rput(1.0,0){\obc}\rput(1.0,0){$\AS\ad$}
\rput(2.4,0){\obc}\rput(2.4,0){$\AS$}
	}
	\def\aobjb{
\psline(0,0)(1.5,0)
\psline{->}(0,0)(0.9,0)
\rput(0,0){\cldot}\rput[b](0,0.2){$a$}
\rput(1.5,0){\opdot}\rput[b](1.5,0.2){$a$}
	}
	\def\bobj{
\psline(0,0)(1,0)\rarr(0.85,0)
\rput(0,0){\obc}\rput(1,0){\opdot}
\rput(0,0){$\psi$}\rput[b](1.0,0.2){$a$}
	}
	\def\bobjtr{
\psline(0,0)(2.4,0)
\rarr(0.85,0)\larr(1.95,0)
\rput(0,0){\obc}\rput(0,0){$\psi$}
\rput(1.4,0){\obc}\rput(1.4,0){$\AS\ad$}
\rput(2.4,0){\cldot}\rput[b](2.4,0.2){$a$}
	}
	\def\bobjab{
\psline(0,0)(1,0)
\larr(0.6,0)
\rput(0,0){\obr{.5}}\rput(0,0){$\AS\ad\psi$}
\rput(1,0){\cldot}\rput[b](1.0,0.2){$a$}
	}
	\def\cobj{
\psline(0,0)(0,2)\psline{>->}(0,0.15)(0,1.85)
\rput(0,0){\cldot}\rput(0,2){\opdot}
\rput(0,1){\obc}\rput(0,1){$P$}
\rput[r](-.2,0){$a$}\rput[r](-.2,2){$a$}
	}
	\def\cobjtr{
\psline(0,0)(1,0)\psline{<-}(0.2,0)(1,0)
\psarc(1,1){1.0}{-90}{0}\psarc{->}(1,1){1.0}{-90}{-40}
\rput[lt](1.75,0.25){$a$}
\rput(1,0){\obc}\rput(1,0){$\AS$}
\rput(0,0){\opdot}\rput[b](0,0.2){$a$}
\psline(0,2)(1,2)\psline{->}(0,2)(0.45,2)
\psarc(1,1){1.0}{0}{90}\psarc{->}(1,1){1.0}{0}{60}
\rput[lb](1.8,1.9){$a$}
\rput(1,2){\obc}\rput(1,2){$\AS\ad$}
\rput(0,2){\cldot}\rput[b](0,2.2){$a$}
\rput(2,1){\obc}\rput(2,1){$P$}
	}
	\def\cobjab{
\psline(0,0)(0,2)\psline{>->}(0,1.85)(0,0.15)
\rput(0,0){\opdot}\rput(0,2){\cldot}
\rput(0,1){\obr{.7}}\rput(0,1){$\AS\ad P\AS$}
\rput[r](-.2,0){$a$}\rput[r](-.2,2){$a$}
	}
	\def\dobj{
\psline(0,0)(0,2)\psline{<->}(0,0.15)(0,1.85)
\rput(0,0){\opdot}\rput(0,2){\opdot}
\rput(0,1){\obc}\rput(0,1){$\Psi$}
\rput[r](-.2,0){$a$}\rput[r](-.2,2){$b$}
	}
	\def\dobjtr{
\psline(2,1)(2,2)\psline{->}(2,1)(2,1.85)
\psline(0,0)(1,0)\psline{->}(0.0,0)(0.4,0)
\psarc(1,1){1.0}{-90}{0}\psarc{<-}(1,1){1.0}{-50}{0}
\rput[lt](1.75,0.25){$a$}
\rput(1,0){\obc}\rput(1,0){$\AS\ad$}
\rput(0,0){\cldot}\rput[b](0,0.2){$a$}
\rput(2,2){\opdot}\rput[r](1.8,2){$b$}
\rput(2,1){\obc}\rput(2,1){$\Psi$}
	}
	\def\dobjab{
\psline(0,0)(0,2)\psline{>->}(0,0.15)(0,1.85)
\rput(0,0){\cldot}\rput(0,2){\opdot}
\rput(0,1){\obr{0.5}}\rput(0,1){$\AS\ad\Psi$}
\rput[r](-.2,0){$a$}\rput[r](-.2,2){$b$}
	}
	\def\eobja{
\psline(0,0)(3.4,0)
\psline{->}(0,0)(0.4,0)\psline{->}(0,0)(3.2,0)
\psline{<-}(1.5,0)(2,0)
\rput[b](1.7,0.2){$b$}
\rput(0,0){\cldot}\rput[b](0,0.2){$a$}
\rput(3.4,0){\opdot}\rput[b](3.4,0.2){$a$}
\rput(1.0,0){\obc}\rput(1.0,0){$\Psi^\mnone$}
\rput(2.4,0){\obc}\rput(2.4,0){$\Psi$}
	}
	\def\eobjb{
\psline(0,0)(1.5,0)
\psline{->}(0,0)(0.9,0)
\rput(0,0){\cldot}\rput[b](0,0.2){$a$}
\rput(1.5,0){\opdot}\rput[b](1.5,0.2){$a$}
	}
	\def\eobjc{
\psline(0,0)(3.4,0)
\psline{->}(0,0)(0.4,0)\psline{->}(0,0)(3.2,0)
\psline{<-}(1.5,0)(2,0)
\rput[b](1.7,0.2){$a$}
\rput(0,0){\cldot}\rput[b](0,0.2){$b$}
\rput(3.4,0){\opdot}\rput[b](3.4,0.2){$b$}
\rput(1.0,0){\obc}\rput(1.0,0){$\Psi^\mnone$}
\rput(2.4,0){\obc}\rput(2.4,0){$\Psi$}
	}
	\def\eobjd{
\psline(0,0)(1.5,0)
\psline{->}(0,0)(0.9,0)
\rput(0,0){\cldot}\rput[b](0,0.2){$b$}
\rput(1.5,0){\opdot}\rput[b](1.5,0.2){$b$}
	}

	\rput(0,0.3){
\rput[l](0,0){(a)}
\rput(1,0){\aobja}
\rput(4.9,0){$=$}
\rput(5.4,0){\aobjb}
	}

	\rput(0,-.2){
\rput[l](0,-1){(b)}
\rput(1.1,-1){\bobj}
\rput(3.0,-1){\bobjtr}
\rput(5.76,-1){$=$}
\rput(6.5,-1){\bobjab}
	}

	\rput(0,-2.5){
\rput[l](0,-1){(c)}
\rput(1.5,-2){\cobj}
\rput(2.7,-2){\cobjtr}
\rput(5.6,-1){$=$}
\rput(6.8,-2){\cobjab}
	}

	\rput(0,-5.5){
\rput[l](0,-1){(d)}
\rput(1.5,-2){\dobj}
\rput(2.7,-2){\dobjtr}
\rput(5.7,-1){$=$}
\rput(6.8,-2){\dobjab}
	}

	\rput(0,-8){
\rput(1,-1){\eobja}
\rput(4.9,-1){$=$}
\rput(5.4,-1){\eobjb}
\rput[l](0,-1.5){(e)}
\rput(1,-2){\eobjc}
\rput(4.9,-2){$=$}
\rput(5.4,-2){\eobjd}
	}

\end{pspicture}
$$
\caption{ (a) The adjoint of a transposer is its inverse. (b)--(d) Transposers
change open to closed nodes, or vice versa.  (e) Inverse of an entangled ket.}
\label{fgr5}
\end{figure}

Let $\{|a_j\rgl\}$ be an orthonormal basis for the space $\HS_a$, and on the
tensor product $\HS_a\ot\HS_a$ of $\HS_a$ with itself
define the \emph{transposer} $\AS$ and its adjoint $\AS\ad$ as a fully
entangled ket and bra,
\begin{equation}
 |\AS\rgl = \sum_j |a_j\rgl\ot|a_j\rgl,\quad 
 \lgl\AS| =  \sum_j \lgl a_j|\ot\lgl a_j|.
\label{eqn5}
\end{equation}
 with normalization
\begin{equation}
 \lgl\AS|\AS\rgl=d_a
\label{eqn6}
\end{equation}
(Readers who prefer to distinguish the two copies of $\HS_a$ may think of $\AS$
as defined on $\HS_a\ot\HS_{a'}$, change the second $a_j$ to $a'_j$ in each of
the expressions in \eqref{eqn5}, and make appropriate modifications in
Fig.~\ref{fgr5}.)

Figure~\ref{fgr5}(a) shows that $\AS\ad$ is the inverse of $\AS$, which makes
$|\AS\rgl$ the analog of a unitary map.  Parts (b) through (d) show how a
transposer can be used to convert open to closed nodes and vice versa.  In part
(b) the transposer applied to a ket $\psi$ yields a bra $\AS\ad\psi$ whose row
vector in the basis $\{|a_j\rgl\}$ is the transpose, in the usual sense of that
term (without complex conjugation), of the column vector corresponding to the
ket in this same basis.  The transpose of an operator $P$, part (c), requires
the use of both $\AS$ and $\AS\ad$, and the result can be denoted $\AS\ad
P\AS$, following the rule given in Sec.~\ref{sct2c}.  The conventional notation
$P\trp$ is more compact, but carries less information, since it does not show
the basis dependence of the resulting operator.

Part (d) of the figure shows how a transposer can change an entangled
ket $\Psi\in\HS_{ab}$ into a map $\AS\ad\Psi\in\LS_{ba}$.  Although such a
change is not ordinarily called a ``transpose,'' the term (or, if one prefers,
``generalized transpose'') seems appropriate.  In Dirac notation it takes a
particularly simple form if, along with the orthonormal basis $\{|a_j\rgl\}$
defining $\AS$, one introduces a basis $\{|b_k\rgl\}$ for $\HS_b$, and expands
both $\Psi$ and $ M=\AS\ad\Psi$ in these bases:
\begin{equation}
 |\Psi\rgl = \sum_{jk}\mu_{jk}|a_j\rgl\ot|b_k\rgl,\quad
 M=\lgl\AS|\Psi\rgl = \sum_{jk}\mu_{jk}|b_k\rgl\lgl a_j|,
\label{eqn7}
\end{equation}
Note that the matrix $\mu_{jk}$ is the same in both expressions; all the
transpose does is change $|a_j\rgl$ to $\lgl a_j|$.  Also note that in the case
$d_a=d_b$, if $|\Psi\rgl$ is a fully-entangled ket normalized so that
$\lgl\Psi|\Psi\rgl=d_a$, $M$ is a unitary map, and vice versa.

The \emph{rank} $\Rn(M)$ of the map $M$ in \eqref{eqn7} can be defined as the
rank of the matrix $\mu_{jk}$ (the dimension of the space spanned by its
columns). This definition depends only on $M$ and not on the choice of bases,
because the rank of a matrix is left unchanged upon left and right
multiplication by nonsingular square matrices: see p.~13 in \cite{HrJh85}.  The
rank of $|\Psi\rgl$ can be defined in exactly the same way, and is often called
the \emph{Schmidt rank}, which we sometimes denote by $\sg$, because it is the
number of positive Schmidt coefficients in the diagonal expansion
\begin{equation}
 |\Psi\rgl = \sum_j \lm_j|\bar a_j\rgl\ot|\bar b_j\rgl,\quad \lm_j\geq 0,
\label{eqn8}
\end{equation}
obtained using appropriate orthonormal bases $\{|\bar a_j\rgl\}$ and $\{|\bar
b_j\rgl\}$.  These Schmidt coefficients are none other than the singular values
of the map $M$.

Continuing the list of useful parallels between maps and entangled kets, if the
Schmidt rank $\sg$ of $|\Psi\rgl$ is equal to $d_a=d_b$, one can define
its \emph{inverse} using the formula
\begin{equation}
 \lgl(\Psi^{-1})\ad| = \sum_{jk}\nu_{kj} \lgl a_j|\ot\lgl b_k|
 = \sum_j \lm_j^{-1}\lgl\bar a_j|\ot\lgl\bar b_j|,
\label{eqn9}
\end{equation}
where $\nu$ is the inverse of the $\mu$ matrix in \eqref{eqn7}:
$\sum_k\mu_{jk}\nu_{kl}=\dl_{jl}$.  As indicated in the diagram in
Fig.~\ref{fgr5}(e), $\Psi^{-1}$ is both the left and the right inverse of
$\Psi$.  (The dagger on the left side of \eqref{eqn9} is absent in the
diagram; see the remarks accompanying Fig.~\ref{fgr1}.) Of course, $\Psi$ is
the inverse of $\Psi^{-1}$.  Although the first expression in \eqref{eqn9}
defines the inverse using a particular basis, it is easy to show that the
resulting bra does not depend on the choice of basis.  In addition, it is worth
noting that the inverse of $\Psi\ad$ is the same as the adjoint of $\Psi^{-1}$,
so the order of minus one and dagger in $\Psi^{-1\dagger}$ does not matter.
When the rank of $|\Psi\rgl$ is smaller than $d_a$, or if $d_a\neq d_b$, one
can define a ``generalized inverse'' by only retaining the terms with $\lm_j>0$
on the right side of \eqref{eqn9}. See the discussion, for maps, in
\cite{HrJh85}, p.~421.  In this paper we do not need such a generalized
inverse, so will not discuss it further.

In the case of an object with 3 or more nodes, such as $Y$ in Fig.~\ref{fgr3},
the discussion of rank becomes more complicated, because $Y$ can be turned into
a map from one space to another in several different ways.  The most natural
interpretation of $Y$, given that the $c$ node is closed and the $a$ and $b$
nodes are open, is as a map $Y_{ab;c}$ from $\HS_c$ to $\HS_{ab}$, and as such
it has a rank which can be written as $\Rn(Y_{ab;c})$.  But one can just as
well regard $Y$ as a map $Y_{bc;a}$ from $\HS_a\ad$ to
$\HS_b\HS_c\ad=\LS_{bc}$, and as such it will have a rank $\Rn(Y_{bc;a})$
different (in general) from $\Rn(Y_{ab;c})$.  Or, equivalently, $\Rn(Y_{bc;a})$
is the rank of $Y'=\AS\ad Y\CS$, obtained by applying transposers to the $a$
and $c$ nodes of $Y$, when it is regarded as a map $Y'_{bc;a}$ from $\HS_a$ to
$\HS_{bc}$.  One way to think about these different possibilities is to write
$Y$ in the Dirac form
\begin{equation}
 Y= \sum_{j,k,l} \Blp\lgl a_j,b_k|Y|c_l\rgl\Brp
   |a_j\rgl\ot|b_k\rgl\ot\lgl c_l|.
\label{eqn10}
\end{equation}
using orthonormal bases $\{|a_j\rgl\}$, $\{|b_k\rgl\}$, $\{|c_l\rgl\}$.  If one
regards $\lgl a_j,b_k|Y|c_l\rgl$ as a matrix with columns labeled by $l$ and
rows by the double label $(j,k)$, its rank will be $\Rn(Y_{ab;c})$, whereas to
obtain $\Rn(Y_{bc;a})$ consider it a matrix with columns labeled by $j$, and
rows by the double label $(k,l)$.  There is no reason to expect these two ranks
to be equal, and of course there is a third possibility, $\Rn(Y_{ac;b})$, which
could be different from the other two.  Note that our subscript notation simply
divides the collection of nodes into two sets, with the object understood as a
map from the space labeled by the second set to that labeled by the first.
Thus $Y_{ab;c}$ and $Y_{ba;c}$ denote the same map.  In addition, since
transposing a matrix does not change its rank, $\Rn(Y_{ab;c})$ is the same as
$\Rn(Y_{c;ab})$, even though the maps are distinct.  Consequently there are
only three possible ranks associated with the object $Y$, and it makes no
difference if one changes open nodes to closed nodes or vice versa using
appropriate transposers.  Atemporal diagrams help one to see that a single
object can be associated with several different ranks, but they do not in and
of themselves place any constraints on these different possibilities.

	\subsection{The virtue of being careless}
\label{sct2f}

The system of diagrams introduced above is precise in the sense that the
(tensor product) Hilbert space identified with an object is specified by the
types, open or closed, of nodes in its diagram, along with the labels attached
to them, with the space $\HS\ad$ of bras carefully distinguished from the space
$\HS$ of kets, as in Dirac notation.  One can convert an open to a closed node
or vice versa by introducing a transposer, Sec.~\ref{sct2e}, but in this case
the symbol for the transposer becomes part of the diagram, as in
Fig.~\ref{fgr5}(b), (c), and (d), and indicates precisely which basis was
employed for the transposition, a concept that is not basis-independent.

While such precision is often valuable, there are circumstances, especially
when one is carrying out a first, informal analysis of a problem, when it is
useful to \emph{ignore} the difference between open and closed nodes, or the
direction of arrows on contractions lines, and treat two diagrams that differ
only in these respects as ``essentially the same.''  To put the matter in a
different way, there are analogies between two diagrams which are ``identical''
in this loose sense; analogies which are worth exploring, because they might
produce helpful insights.

For example, if $\Psi$ in Fig.~\ref{fgr5}(d) is a fully-entangled ket on two
Hilbert spaces of the same dimension, the corresponding map $\AS\ad\Psi$ is a
unitary map, up to multiplication by a constant.  Thus ``fully-entangled'' and
``unitary'' are ``essentially the same.'' This helps one understand, at an
intuitive level, the essence of teleportation as discussed below in
Sec.~\ref{sct3}, using the diagrams in Fig.~\ref{fgr7}, in which (c) and (d)
both represent unitary channels.  Another example is Schmidt rank, see the
discussion in connection with \eqref{eqn8}, which is the ordinary rank of a map
which is ``essentially the same'' as an entangled ket.  The Kraus rank
discussed below in Sec.~\ref{sct5a}  can in a similar way be identified with
the ordinary rank of the ``cross operator'' $V$ in Fig.~\ref{fgr10}(b), which
maps the $f$ node part into the $a,b$ node part, without worrying about whether
these nodes are open or closed; see the discussion of rank in Sec.~\ref{sct2e}.
A fourth example, not illustrated in the present paper, is that of letting time
flow backwards in a quantum circuit.

In some cases one can easily identify the formal equivalence that makes such
diagrams ``essentially the same.''  Thus a fully-entangled state has equal
Schmidt coefficients, and these correspond to the equal singular values that
characterize a scalar multiple of a unitary operator.  However, experience
suggests that it is generally more fruitful to think about the analogy, and ask
whether it provides some useful insight, than it is to work out the formal
correspondence.

	\section{Interchange; Teleportation and Dense Coding}
\label{sct3}

	\subsection{Interchange}
\label{sct3a}

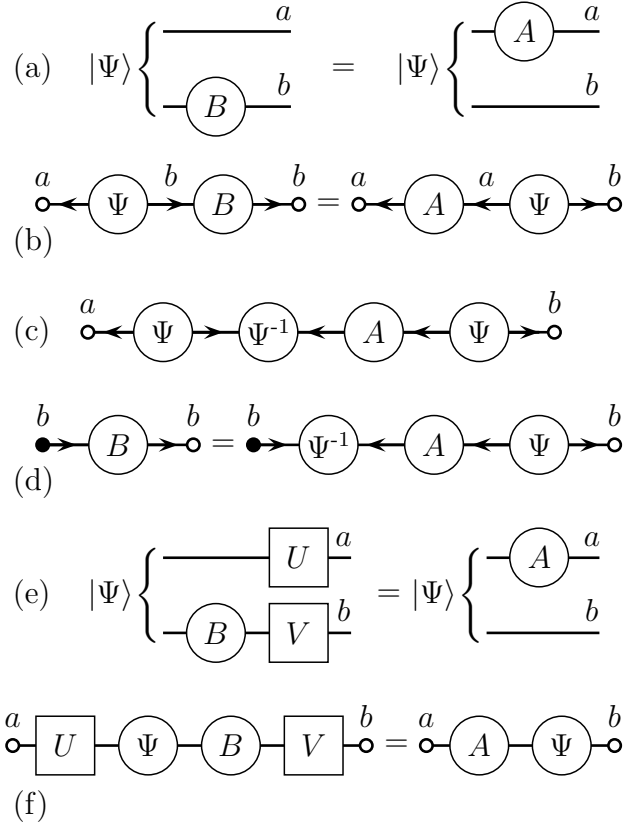
\begin{figure}[h]
$$
\begin{pspicture}(0,-10.5)(8.0,0.3) 
\newpsobject{showgrid}{psgrid}{subgriddiv=1,griddots=10,gridlabels=6pt}
\psset{
labelsep=2.0,
arrowsize=0.150 1,linewidth=\lwd}
\def\lbrk{\left\{\vvv{0.5}\right.}
	\def\hedd{\mbox{
\parbox[t]{2cm}{\centering Original\\[-.5ex] object}%
\parbox[t]{3.7cm}{\centering Transposed object}
\parbox[t]{1.8cm}{\centering Short\\[-.5ex] notation}
	}}
	\def\aobja{
\psline(0,1)(1.7,1)
\psline(0,0)(1.7,0)
\rput[r](0,0.5){$\lbrk$}
\rput[r](-0.4,0.5){$|\Psi\rgl$}
\rput(0.7,1){\obc}\rput(0.7,1){$A$}
\rput[b](1.6,1.15){$a$}
\rput[b](1.6,0.15){$b$}
	}
	\def\aobjb{
\psline(0,1)(1.7,1)
\psline(0,0)(1.7,0)
\rput[r](0,0.5){$\lbrk$}
\rput[r](-0.4,0.5){$|\Psi\rgl$}
\rput(0.7,0){\obc}\rput(0.7,0){$B$}
\rput[b](1.6,1.15){$a$}
\rput[b](1.6,0.15){$b$}
	}
	\def\bobja{
\psline(0,0)(3.4,0)
\psline{<->}(0.2,0)(3.2,0)
\psline{<-}(1.5,0)(2.5,0)\rput[b](1.7,0.25){$a$}
\rput(0,0){\opdot}\rput[b](0,0.25){$a$}
\rput(3.4,0){\opdot}\rput[b](3.4,0.25){$b$}
\rput(1,0){\obc}\rput(1,0){$A$}
\rput(2.4,0){\obc}\rput(2.4,0){$\Psi$}
	}
	\def\bobjb{
\psline(0,0)(3.4,0)
\psline{<->}(0.2,0)(3.2,0)
\psline{->}(1.0,0)(1.9,0)\rput[b](1.7,0.25){$b$}
\rput(0,0){\opdot}\rput[b](0,0.25){$a$}
\rput(3.4,0){\opdot}\rput[b](3.4,0.25){$b$}
\rput(1,0){\obc}\rput(1,0){$\Psi$}
\rput(2.4,0){\obc}\rput(2.4,0){$B$}
	}
	\def\cobj{
\psline(0,0)(6.2,0)
\psline{<->}(0.2,0)(6.0,0)
\psline{->}(0,0)(1.8,0)
\psline{<-}(2.9,0)(6.0,0)\psline{<-}(4.3,0)(6.0,0)
\rput(0,0){\opdot}\rput[b](0,0.25){$a$}
\rput(6.2,0){\opdot}\rput[b](6.2,0.25){$b$}
\rput(1,0){\obc}\rput(1,0){$\Psi$}
\rput(2.4,0){\obc}\rput(2.4,0){$\Psi^{\hbox{\scriptsize-1}}$}
\rput(3.8,0){\obc}\rput(3.8,0){$A$}
\rput(5.2,0){\obc}\rput(5.2,0){$\Psi$}
	}
	\def\dobja{
\psline(0,0)(4.8,0)
\psline{->}(0,0)(0.45,0)
\psline{<-}(1.5,0)(4,0)
\psline{<->}(2.9,0)(4.6,0)
\rput(0,0){\cldot}\rput[b](0,0.25){$b$}
\rput(4.8,0){\opdot}\rput[b](4.8,0.25){$b$}
\rput(1,0){\obc}\rput(1,0){$\Psi^{\hbox{\scriptsize-1}}$}
\rput(2.4,0){\obc}\rput(2.4,0){$A$}
\rput(3.8,0){\obc}\rput(3.8,0){$\Psi$}
	}
	\def\dobjb{
\psline(0,0)(2,0)
\psline{->}(0,0)(0.45,0)
\psline{->}(0,0)(1.8,0)
\rput(0,0){\cldot}\rput[b](0,0.25){$b$}
\rput(2,0){\opdot}\rput[b](2,0.25){$b$}
\rput(1,0){\obc}\rput(1,0){$B$}
	}
	\def\eobja{
\psline(0,1)(1.5,1)
\psline(0,0)(1.5,0)
\rput[r](0,0.5){$\lbrk$}
\rput[r](-0.4,0.5){$|\Psi\rgl$}
\rput(0.7,1){\obc}\rput(0.7,1){$A$}
\rput[b](1.4,1.15){$a$}
\rput[b](1.4,0.15){$b$}
	}
	\def\eobjb{
\psline(0,1)(2.5,1)
\psline(0,0)(2.5,0)
\rput[r](0,0.5){$\lbrk$}
\rput[r](-0.4,0.5){$|\Psi\rgl$}
\rput(0.7,0){\obc}\rput(0.7,0){$B$}
\rput(1.8,0){\obx}\rput(1.8,0){$V$}
\rput(1.8,1){\obx}\rput(1.8,1){$U$}
\rput[b](2.4,1.15){$a$}
\rput[b](2.4,0.15){$b$}
	}
	\def\fobja{
\psline(0,0)(2.5,0)
\rput(0,0){\opdot}\rput[b](0,0.25){$a$}
\rput(0.7,0){\obc}\rput(0.7,0){$A$}
\rput(1.8,0){\obc}\rput(1.8,0){$\Psi$}
\rput(2.5,0){\opdot}\rput[b](2.5,0.25){$b$}
	}
	\def\fobjb{
\psline(0,0)(4.7,0)
\rput(0,0){\opdot}\rput[b](0,0.25){$a$}
\rput(0.7,0){\obx}\rput(0.7,0){$U$}
\rput(1.8,0){\obc}\rput(1.8,0){$\Psi$}
\rput(2.9,0){\obc}\rput(2.9,0){$B$}
\rput(4.0,0){\obx}\rput(4.0,0){$V$}
\rput(4.7,0){\opdot}\rput[b](4.7,0.25){$b$}
	}
\rput[l](0,-0.5){(a)}
\rput(2.0,-1){\aobjb}
\rput(4.4,-0.5){$=$}
\rput(6.1,-1){\aobja}

\rput[l](0,-2.8){(b)}
\rput(0.4,-2.3){\bobjb}
\rput(4.2,-2.3){$=$}
\rput(4.6,-2.3){\bobja}

\rput[l](0,-4){(c)}
\rput(1.0,-4){\cobj}

\rput[l](0,-6){(d)}
\rput(0.4,-5.5){\dobjb}
\rput(2.8,-5.5){$=$}
\rput(3.2,-5.5){\dobja}

\rput[l](0,-7.5){(e)}
\rput(2.0,-8){\eobjb}
\rput(5.0,-7.5){$=$}
\rput(6.3,-8){\eobja}

\rput[l](0,-10.3){(f)}
\rput(0,-9.5){\fobjb}
\rput(5.1,-9.5){$=$}
\rput(5.5,-9.5){\fobja}
\end{pspicture}
$$
\caption{ The circuit equation (a), or its atemporal equivalent (b), for the
unknown operator $B$ can be solved as shown in (c) and (d) if $\Psi$ has an
inverse. Solving (e) or (f) for $B$ and the unitaries $U$ and $V$ is discussed
in the text.}
\label{fgr6}
\end{figure}

As a first example of how a quantum circuit can be represented by an atemporal
diagram, consider the circuit equation in Fig.~\ref{fgr6}(a): Given an
entangled state $|\Psi\rgl$ on $\HS_a\ot\HS_b$, with $d_b=d_a=d$ and an
operator $A$ on $\HS_a$, find, if possible, an operator $B$ on $\HS_b$ such
that
\begin{equation}
 (I_a\ot B)|\Psi\rgl = (A\ot I_b)|\Psi\rgl.
\label{eqn11}
\end{equation}
Part (b) of the figure shows the equation reduced to atemporal diagrams, and
part (c) shows a strategy for solving it if $|\Psi\rgl$ is of rank $d$, and
thus has an inverse: insert the identity in the form $\Psi^{-1}\Psi$ between
the $a$ node and $A$ on the right side of (b), and it is evident that the final
set of three objects in (c) represent the solution, as written in diagrammatic
form in (d).  It is not very easy to interpret the Dirac equivalent
\begin{equation}
 B = \lgl(\Psi^{-1})\ad| A |\Psi\rgl
\label{eqn12}
\end{equation}
apart from saying it means the same thing as (d) in the figure.

When the rank of $|\Psi\rgl$ is less than $d$, this strategy will not work, and
in general \eqref{eqn11} has no solution.  However, a weaker version,
Fig.~\ref{fgr6}(e), where the problem is to find $B$ and two unitary operators
$U$ and $V$ such that, for a given $|\Psi\rgl$ and $A$,
\begin{equation}
  (U\ot V)(I_a\ot B)|\Psi\rgl = (A\ot I_b)|\Psi\rgl,
\label{eqn13}
\end{equation}
can always be solved, even when one imposes the additional condition that $B$
be unitarily equivalent to $A$, in the sense that there is a unitary map $S$
from $\HS_a$ to $\HS_b$ such that $B=SAS\ad$.  The strategy, see App.~A of
\cite{LoPp01}, is to choose orthonormal bases in which $|\Psi\rgl$ has Schmidt
form \eqref{eqn8}, and define
\begin{equation}
 B=\sum_{jk} \lgl\bar a_j|A|\bar a_k\rgl\cdot |\bar b_j\rgl\lgl\bar b_k|,
\label{eqn14}
\end{equation}
which is obviously unitarily equivalent to $A$ in the sense just defined:
$S=\sum_j|\bar b_j\rgl\lgl\bar a_j|$.  Given this definition for $B$, it is
evident that the two entangled kets
\begin{equation}
 |\Psi_A\rgl = (A\ot I_b)|\Psi\rgl,\quad
 |\Psi_B\rgl = (I_a\ot B)|\Psi\rgl
\label{eqn15}
\end{equation}
are mapped into each other by interchanging the two spaces $\HS_a$ and $\HS_b$,
a symmetry which guarantees that they possess the same set of coefficients when
each is expanded in Schmidt form, relative to appropriate Schmidt bases.
The ``local unitaries'' $U$ and $V$ are then those that map the Schmidt bases
for $|\Psi_B\rgl$ to those of $|\Psi_A\rgl$.

The corresponding atemporal diagram in Fig.~\ref{fgr6}(f), while not used in
deriving this result, immediately suggests an extension, especially with
arrows omitted (the reader can easily supply them): Interchanging two
objects $A$ and $\Psi$ is possible at a cost of unitary equivalence and
introducing additional unitaries.  The objects could be two operators, or an
entangled ket and an entangled bra.

	\subsection{Teleportation and dense coding}
\label{sct3b}

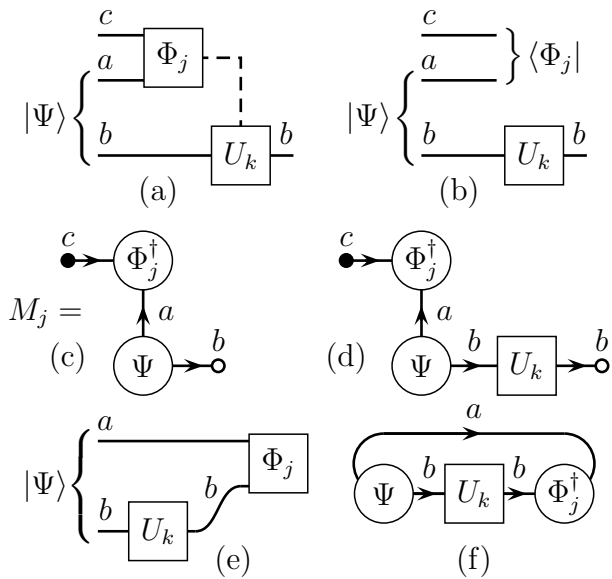
\begin{figure}[h]
$$
\begin{pspicture}(0,-7.7)(8.0,0.1) 
\newpsobject{showgrid}{psgrid}{subgriddiv=1,griddots=10,gridlabels=6pt}
\psset{
labelsep=2.0,
arrowsize=0.150 1,linewidth=\lwd}
\def\lbrk{\left\{\vvv{0.5}\right.}
\def\lbrkb{\left\{\vvv{0.6}\right.}
\def\rbrk{\left.\vvv{0.3}\right\}}
\def\dput(#1)#2#3{\rput(#1){#2}\rput(#1){#3}} 
\def\obcs{\pscircle[fillcolor=white,fillstyle=solid,linewidth=\lwn](0,0){0.35}}
\def\clnr(#1)#2{\dput(#1){\cldot}{\rput[r](-0.2,0){$#2$}}} 
\def\clnl(#1)#2{\dput(#1){\cldot}{\rput[l](0.2,0){$#2$}}} 
\def\clnb(#1)#2{\dput(#1){\cldot}{\rput[b](0,0.2){$#2$}}} 
\def\clnt(#1)#2{\dput(#1){\cldot}{\rput[t](0,-0.2){$#2$}}} 
\def\opnr(#1)#2{\dput(#1){\opdot}{\rput[r](-0.2,0){$#2$}}} 
\def\opnl(#1)#2{\dput(#1){\opdot}{\rput[l](0.2,0){$#2$}}} 
\def\opnb(#1)#2{\dput(#1){\opdot}{\rput[b](0,0.2){$#2$}}} 
\def\opnt(#1)#2{\dput(#1){\opdot}{\rput[t](0,-0.2){$#2$}}} 
\def\rarr(#1){\rput(#1){\psline{->}(-0.2,0)(0,0)}}
\def\larr(#1){\rput(#1){\psline{->}(0.2,0)(0,0)}}
\def\uarr(#1){\rput(#1){\psline{->}(0,-0.2)(0,0)}}
\def\darr(#1){\rput(#1){\psline{->}(0,0.2)(0,0)}}
	\def\hedd{\mbox{
\parbox[t]{2cm}{\centering Original\\[-.5ex] object}%
\parbox[t]{3.7cm}{\centering Transposed object}
\parbox[t]{1.8cm}{\centering Short\\[-.5ex] notation}
	}}
	\def\aobj{
\psline(0,1.6)(1,1.6)
\rput[b](0.1,1.75){$c$}
\psline(0,1)(1,1)
\rput[b](0.1,1.15){$a$}
\psline(0,0)(2.6,0)
\rput[b](0.1,0.15){$b$}
\psline[linestyle=dashed](1,1.3)(1.9,1.3)(1.9,0)
\rput(1,1.3){\obx}\rput(1,1.3){$\Phi_j$}
\rput(1.9,0){\obx}\rput(1.9,0){$U_k$}
\rput[r](0,0.5){$\lbrk$}
\rput[r](-0.4,0.5){$|\Psi\rgl$}
\rput[b](2.5,0.15){$b$}
	}
	\def\bobj{
\psline(0,1.6)(1,1.6)
\rput[b](0.1,1.75){$c$}
\psline(0,1)(1,1)
\rput[b](0.1,1.15){$a$}
\rput(1.2,1.3){$\rbrk$}\rput[l](1.4,1.3){$\lgl\Phi_j|$}
\psline(0,0)(2.2,0)
\rput[b](0.1,0.15){$b$}
\rput(1.5,0){\obx}\rput(1.5,0){$U_k$}
\rput[r](0,0.5){$\lbrk$}
\rput[r](-0.4,0.5){$|\Psi\rgl$}
\rput[b](2.1,0.15){$b$}
	}
	\def\cobj{
\psline(0,1.4)(1,1.4)(1,0)(2,0)
\clnb(0,1.4){c} \opnb(2,0){b}
\dput(1,1.4){\obc}{$\Phi_j\ad$}
\dput(1,0){\obc}{$\Psi$}
\rput[l](1.2,0.7){$a$}
\rarr(0.45,1.4)
\uarr(1,0.8)
\rarr(1.8,0)
	}
	\def\dobj{
\psline(0,1.4)(1,1.4)(1,0)(3.4,0)
\clnb(0,1.4){c} \opnb(3.4,0){b}
\dput(1,1.4){\obc}{$\Phi_j\ad$}
\dput(1,0){\obc}{$\Psi$}
\dput(2.4,0){\obx}{$U_k$}
\rput[l](1.2,0.7){$a$}
\rput[b](1.7,0.2){$b$}
\rarr(0.45,1.4)
\uarr(1,0.8)
\rarr(1.8,0)
\rarr(3.2,0)
	}
	\def\eobj{
\psline(0,1.2)(2.4,1.2)
\rput[b](0.1,1.35){$a$}
\psline(0,0)(1.3,0)
\rput[b](0.1,0.15){$b$}
\psline(1.9,0.6)(2.4,0.6)
\psbezier(1.3,0)(1.6,0)(1.6,0.6)(1.9,0.6)
\rput[b](1.5,0.45){$b$}
\rput(0.8,0){\obx}\rput(0.8,0){$U_k$}
\rput(2.4,0.9){\obx}\rput(2.4,0.9){$\Phi_j$}
\rput[r](0,0.6){$\lbrkb$}
\rput[r](-0.4,0.6){$|\Psi\rgl$}
	}
	\def\fobj{
\psline(0.8,0.8)(3.2,0.8)
\psline{->}(0.8,0.8)(2.1,0.8)
\psline(0.8,0)(3.2,0)
\psline{->}(0.8,0)(1.5,0)
\psline{->}(0.8,0)(2.7,0)
\rput[b](2.0,1.0){$a$}
\rput[b](1.4,0.2){$b$}
\rput[b](2.6,0.2){$b$}
\psarc(0.8,0.4){0.4}{90}{270}
\psarc(3.2,0.4){0.4}{-90}{90}
\rput(3.2,0){\obc}\rput(3.2,0){$\Phi_j\ad$}
\rput(0.8,0){\obc}\rput(0.8,0){$\Psi$}
\rput(2.0,0){\obx}\rput(2.0,0){$U_k$}
	}

\rput(1.2,-2){\aobj}
\rput[B](2,-2.6){(a)}
\rput(5.5,-2){\bobj}
\rput[B](6,-2.6){(b)}

	\rput(0,0.2){
\rput[l](0,-4.3){$M_j=$}
\rput(0.8,-5){\cobj}
\rput[B](0.8,-5){(c)}
\rput(4.5,-5){\dobj}
\rput[B](4.5,-5){(d)}
	}

\rput(1.2,-7){\eobj}
\rput[B](3,-7.5){(e)}
\rput(4.2,-6.5){\fobj}
\rput[B](6.2,-7.5){(f)}

\end{pspicture}
$$
\caption{Circuit diagrams (a) and (b) for teleportation, along with
atemporal diagrams (c) representing $M_j$, and (d) the effective quantum
channel from $\HS_c$ to $\HS_b$.  The dense coding quantum circuit (e) leads to
the atemporal diagram (f).}
\label{fgr7}
\end{figure}

The standard teleportation arrangement shown in Fig.~\ref{fgr7}(a) employs
three Hilbert spaces: $\HS_a$, $\HS_b$ and $\HS_c$, of identical dimension
$d$. An initial fully-entangled state $|\Psi\rgl$ on $\HS_{ab}$ is used to
teleport a state $|c\rgl$ of $\HS_c$, by means of a measurement on $\HS_{ac}$
in a basis $\{|\Phi_j\rgl\}$ of fully-entangled states. The outcome $j$ is used
to select one of a set $\{U_k\}$ of $d^2$ unitary operators to apply to $\HS_b$
in order to yield a final state equal to $V|c\rgl$, where $V$ is some unitary
map, fixed and independent of $j$, from $\HS_c$ to $\HS_a$.  One usually thinks
of $V$ as the identity operator, assuming that an appropriate correspondence
has been set up between these two Hilbert spaces; for a careful discussion see
Sec.~V of \cite{NlCv97}.

In constructing an atemporal diagram to represent teleportation in the form
just discussed, we first ignore the ``classical communication'' step indicated
by the dashed curve in part (a) of the figure---we shall return to this aspect
later---and interpret the measurement outcome as indicating the prior state of
the quantum system, as indicated in the circuit in part (b). (Although not yet
found in elementary textbooks, this is a perfectly consistent way of viewing
measurement outcomes; see Ch.~17 of \cite{Grff02}.)  If we expand the initial
state
\begin{equation}
 |\Om\rgl = |c\rgl\ot|\Psi\rgl = \sum_j |\Phi_j\rgl\ot M_j|c\rgl,
\label{eqn16}
\end{equation}
in the basis $\{|\Phi_j\rgl\}$, the Kraus operator $M_j$ can be formally
expressed in Dirac notation as
\begin{equation}
 M_j = \lgl\Phi_j | \Psi\rgl,
\label{eqn17}
\end{equation}
or perhaps with greater clarity using the atemporal diagram of
Fig.~\ref{fgr7}(c). When the final unitary $U_k$ is added to the diagram,
we arrive at part (d), which provides an atemporal representation
of the quantum channel starting at the initial $c$ and ending at the final $b$
of parts (a) or (b) of the figure.

In interpreting the atemporal diagram in Fig.~\ref{fgr7}(c) it is helpful to
adopt the normalization
\begin{equation}
  \lgl\Psi|\Psi\rgl =  \lgl\Phi_j|\Phi_j\rgl = d,
\label{eqn18}
\end{equation}
 in which case $M_j$ will be a unitary operator.  That is, there is a unitary
map from the $c$ input to the $b$ line preceding $U_k$ in part (a) of the
figure, a map which depends on the measurement outcome $j$.  Note that one does
not have to think of node $b$ in (c) as referring to a time following that at
which the measurement on $\HS_{ac}$ occurs; the channel, in the sense of
statistical correlations within an appropriate framework of histories, is
present both at earlier and later times.  One advantage of an \emph{atemporal}
diagram is that one reaches this conclusion without becoming entangled in
misleading notions like ``wave function collapse.''

Once the unitary nature of (c) has been understood, the role of the final
unitary in (d) becomes obvious: in order to obtain a perfect quantum channel
corresponding to the identity map, the final unitary should undo the action of
the preceding unitary, so that in the case of outcome $j$, one should employ
\begin{equation}
 U_j= M_j\ad.
\label{eqn19}
\end{equation}
This means, of course, using a different unitary for each measurement outcome
$j$, and that is why it is essential in actual teleportation protocols that its
value be transmitted from the point where the measurement occurs to the point
where the unitary correction will occur.  While this ``classical
communication'' plays a critical role in the protocol, it does not play an
essential role in understanding what is going on in quantum mechanical terms.
The virtue of the atemporal diagram is that it focuses attention on the
latter, which is to say the quantum correlations whose correct calculation
requires the consistent use of quantum theory.  Once these correlations are
taken care of, other aspects of the situation can be understood through
straightforward application of ideas from classical physics.

The quantum circuit for dense coding, Fig.~\ref{fgr7}(e), has been drawn in a
way to emphasize the similarity with teleportation.  Initially there is a fully
entangled state $|\Psi\rgl$ on $\HS_{ab}$, and one of an appropriate
collection of $d^2$ unitaries is applied to the $\HS_b$ part, following which a
measurement is carried out in a fully-entangled basis $\{|\Phi_j\rgl\}$.  (The
line connecting $U_k$ to $\Phi_j$ is drawn curved in (e) as a reminder that one
typically thinks of this process as transmitting information from one physical
location to another, but this is obviously not essential for understanding the
quantum physics of the situation.)  Upon assuming, as before, that the
measurement outcome represents the prior quantum state, one arrives at the
atemporal diagram in (f).  (The reader may want to insert the obvious analog of
part (b) of the figure as an intermediate step.)

This closed loop, as it contains no nodes, is a complex number whose absolute
square can be interpreted as a probability up to a suitable normalization
constant. If, as before, the normalization \eqref{eqn18} is adopted, so that
$\Phi_j\ad\Psi$ is a unitary operator, dividing by $d^2$ will yield the
conditional probability of measurement outcome $j$ given the use of the unitary
$U_k$.  The conventional choice which makes the loop equal to 0 for $j\neq k$
is obtained using \eqref{eqn19}.  Since this is an atemporal diagram, one could
just as well have drawn (f) with $\Phi_j\ad$ to the left of $\Psi$, whence it
is evident that (f) is obtained from (d) by ``closing the loop.''  Whereas the
close connection of teleportation with dense coding has been pointed out in the
past, see in particular \cite{Wrnr01}, the diagrammatic approach provides a
particularly simple way of seeing the relationship.

	\section{Unambiguous (Conclusive) Teleportation}
\label{sct4}

	\subsection{Introduction}
\label{sct4a}

If the state $|\Psi\rgl$ in Fig.~\ref{fgr7} is an entangled state of Schmidt
rank $d=d_a=d_b=d_c$ but not fully entangled, ordinary teleportation of some
arbitrary state is not possible, but a modified protocol, which we call
\emph{unambiguous} teleportation, will succeed with a probability less than 1.
In one version, Alice carries out a POVM measurement on the initial state
$|\Om\rgl$ in \eqref{eqn16} using a collection of positive operators $\{G_j,\,
j=0,1,\ldots\}$ on $\HS_{ca}$ which sum to the identity. If the outcome
corresponds to $G_j$ for some $j\geq 1$, the \emph{conclusive} case, Bob
applies an appropriate unitary correction $U_j$, and the initial state $|c\rgl$
on $\HS_c$ has been successfully teleported to $\HS_b$.  However, if it
corresponds to $G_0$, the \emph{inconclusive} case, the attempted teleportation
has failed in the sense that in general there is no way Bob can apply a
correction which will ensure the correct output. In a different version, Alice
measures $|\Om\rgl$ in a fully-entangled basis of $\HS_{ca}$, as in the
standard protocol, but Bob upon learning the value of $j$ attempts to carry out
a nonunitary correction $L_j$, which, if successful, completes the
teleportation.  The two can be combined: some general POVM by Alice and a
nonunitary correction by Bob when he learns the outcome, as discussed below in
Sec.~\ref{sct4c}

The adjective ``unambiguous'' refers to the fact that at least one of the
parties carrying out the process, who can always inform the other, is aware of
whether the protocol succeeds (conclusive case) or fails (inconclusive case).
While the term ``conclusive teleportation'' is often employed, we think
``unambiguous'' has a somewhat clearer meaning. Also, its use in this context
agrees with the closely-related idea of unambiguous discrimination---see, for
example, \cite{ChBr98}---and it is more precise than the term ``conclusive'' as
used at present by Mor (the originator of ``conclusive teleportation'') and his
associates: see Sec.~\ref{sct4e}. (Other names are sometimes employed, e.g.,
probabilistic teleportation.)
For any protocol of this type, let $q_j$ be the probability that Alice obtains
outcome $j$, and $r_j$ for $j\geq 1$ the probability that Bob's correction will
succeed---if the latter is a unitary operation, $r_j=1$.  Then the overall
probability of success is
\begin{equation}
  p_s = \sum_{j\geq 1} p_j = \sum_{j\geq 1} q_j r_j,
\label{eqn20}
\end{equation}
with $p_j=q_j r_j$ the probability that Alice's outcome is $j$ \emph{and} Bob's
correction is successful.

What might at first seem like a completely different strategy is for Alice to
carry out \emph{unambiguous entanglement concentration} on $|\Psi\rgl$ to
obtain a fully-entangled state $|\Psi_f\rgl$.  If the process is successful,
teleportation can then be carried out using the standard protocol, employing
$|\Psi_f\rgl$ as the entangled resource.  The optimum probability for
successful unambiguous entanglement concentration is $p_c=d\lm_m^2$,
Sec.~\ref{sct4b}, where $\lm_m$ is the minimum Schmidt coefficient in
\eqref{eqn8}, and consequently it is always possible to carry out unambiguous
teleportation using a two-step process with this probability of overall
success. On the other hand, $p_c$ is also an upper bound on the probability of
success of \emph{any} unambiguous teleportation protocol for, as pointed out on
p.~90 of \cite{BrHM04}, Alice can generate a fully-entangled pair in her
laboratory and teleport half of it to Bob, thus effecting unambiguous
entanglement concentration by means of unambiguous teleportation.  Of course,
the unambiguous entanglement concentration could equally well be carried out by
Bob, with the same probability of success.

Based on this idea one can construct various protocols of the type mentioned
earlier that have the maximum probability of success allowed by the
partially-entangled state used as a resource.  Some of these are considered in
Sec.~\ref{sct4d} following an analysis in Sec.~\ref{sct4c} of the general case
of an arbitrary POVM by Alice leading to an arbitrary (in general nonunitary)
correction by Bob.  We have found this analysis helpful in, among other things,
understanding why it is that in the case of unambiguous teleportation, just as
in regular teleportation, Alice and Bob learn nothing about the nature of the
teleported state, despite the fact that in a general protocol of the sort
described in \ref{sct4c}, Alice \emph{can} acquire statistical information
about the initial state.

While our results could no doubt be obtained without using atemporal diagrams,
we think they add both motivation and clarity to arguments which are more
difficult to understand when expressed in purely algebraic form.  Our approach
serves to provide a unified perspective on a number of different results in the
literature, as discussed in Sec.~\ref{sct4e}.

	\subsection{Entanglement concentration}
\label{sct4b}

\begin{figure}[h]
$$
\begin{pspicture}(0,-1.4)(8.0,0.5) 
\newpsobject{showgrid}{psgrid}{subgriddiv=1,griddots=10,gridlabels=6pt}
\psset{
labelsep=2.0,
arrowsize=0.150 1,linewidth=\lwd}
\def\dput(#1)#2#3{\rput(#1){#2}\rput(#1){#3}} 
\def\lbrk{\left\{\vvv{0.5}\right.}
\def\lbrkb{\left\{\vvv{0.6}\right.}
\def\obcs{\pscircle[fillcolor=white,fillstyle=solid,linewidth=\lwn](0,0){0.35}}
\def\rbrk{\left.\vvv{0.6}\right\}}
\def\clnr(#1)#2{\dput(#1){\cldot}{\rput[r](-0.2,0){$#2$}}} 
\def\clnl(#1)#2{\dput(#1){\cldot}{\rput[l](0.2,0){$#2$}}} 
\def\clnb(#1)#2{\dput(#1){\cldot}{\rput[b](0,0.2){$#2$}}} 
\def\clnt(#1)#2{\dput(#1){\cldot}{\rput[t](0,-0.2){$#2$}}} 
\def\opnr(#1)#2{\dput(#1){\opdot}{\rput[r](-0.2,0){$#2$}}} 
\def\opnl(#1)#2{\dput(#1){\opdot}{\rput[l](0.2,0){$#2$}}} 
\def\opnb(#1)#2{\dput(#1){\opdot}{\rput[b](0,0.2){$#2$}}} 
\def\opnt(#1)#2{\dput(#1){\opdot}{\rput[t](0,-0.2){$#2$}}} 
\def\rarr(#1){\rput(#1){\psline{->}(-0.2,0)(0,0)}}
\def\larr(#1){\rput(#1){\psline{->}(0.2,0)(0,0)}}
\def\uarr(#1){\rput(#1){\psline{->}(0,-0.2)(0,0)}}
\def\darr(#1){\rput(#1){\psline{->}(0,0.2)(0,0)}}
	\def\aobja{
\psline(0,0)(2,0)
\dput(1,0){\obc}{$\Psi_f$}
\larr(0.2,0)\rarr(1.8,0)
\opnb(0,0){a}
\opnb(2,0){b}
	}
	\def\aobjb{
\psline(0,0)(3.4,0)
\dput(1,0){\obc}{$K$}
\dput(2.4,0){\obc}{$\Psi$}
\larr(0.2,0)
\larr(1.55,0)
\rput[b](1.7,0.2){$a$}
\rarr(3.2,0)
\opnb(0,0){a}
\opnb(3.4,0){b}
	}
	\def\bobja{
\psline(0,0)(2,0)
\dput(1,0){\obc}{$K$}
\larr(0.2,0)\larr(1.55,0)
\opnb(0,0){a}
\clnb(2,0){a}
	}
	\def\bobjb{
\psline(0,0)(3.4,0)
\dput(2.4,0){\obc}{$\Psi^{\hbox{\scriptsize-1}}$}
\dput(1,0){\obc}{$\Psi_f$}
\larr(0.2,0)
\rarr(1.8,0)
\rput[b](1.7,0.2){$b$}
\larr(2.95,0)
\opnb(0,0){a}
\clnb(3.4,0){a}
	}

\rput[l](0,0){(a)}
\rput(1.1,0){$\mu$}
\rput(1.5,0){\aobja}
\rput(4.0,0){$=$}
\rput(4.4,0){\aobjb}

\rput[l](0,-1){(b)}
\rput(1.0,-1){\bobja}
\rput(3.5,-1){$=$}
\rput(4.0,-1){$\mu$}
\rput(4.4,-1){\bobjb}

\end{pspicture}
$$
\caption{Definition of operator $K$ using $\Psi^{-1}$.}
\label{fgr8}
\end{figure}
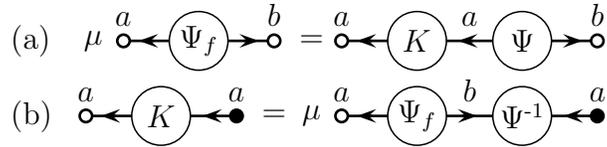

The essence of ordinary teleportation resides in the observation that the
operator $M_j$ in \eqref{eqn17}, with diagram in Fig.~\ref{fgr7}(c), is for
every $j$ a multiple of a unitary operator.  This is no longer the case if
$|\Psi\rgl$ is not fully entangled, but if it has a Schmidt rank of $d$, and
thus an inverse in the sense discussed in Sec.~\ref{sct2e}, unitarity can be
restored by inserting a suitable operator $K$ in the $a$ link in part (c) of
Fig.~\ref{fgr7}, provided $K|\Psi\rgl$ is a constant $\mu$ times a fully
entangled state $|\Psi_f\rgl$, as indicated in Fig.~\ref{fgr8}(a).  The
explicit form of $K$ can be obtained using the inverse $\Psi^{-1}$ of $\Psi$,
as shown in Fig.~\ref{fgr8}(b).  In particular, it is convenient to assume that
\begin{equation}
 |\Psi_f\rgl=(1/\sqrt{d}\,)\sum_j |\bar a_j\rgl\ot|\bar b_j\rgl,
\label{eqn21}
\end{equation}
using the same bases as in \eqref{eqn8}, in which case
\begin{equation}
  K=(\mu/\sqrt{d}\,)\sum_j(1/\lm_j)|\bar a_j\rgl\lgl\bar a_j|.
\label{eqn22}
\end{equation}

Unambiguous entanglement concentration means that Alice carries out the
operation $K$ in an unambiguous manner, using an apparatus which clearly
indicates success or failure; e.g., think of a green or red light going on.
Further details and some subtleties are considered in App.~A. The maximum
probability of success (the green light going on), which is also achievable, is
given by
\begin{equation}
  p_c = \lgl\Psi|K\ad K|\Psi\rgl = d\lm_m^2
\label{eqn23}
\end{equation}
when $\mu=\sqrt{d}\,\lm_m$ is chosen so that the maximum eigenvalue of $K\ad K$
equal to 1; here $\lm_m$ is the minimum of the coefficients in the Schmidt
expansion \eqref{eqn8} for $|\Psi\rgl$. (Note that in the special case in which
$|\Psi\rgl$ is fully entangled, $\lm_m=1/\sqrt{d}$ and the probability of
success is $p_c=1$, as expected.)
Needless to say, the same thing can be achieved by placing $\Psi^{-1}$ in the
$b$ rather than in the $a$ link of Fig.~\ref{fgr7}(c); i.e., Bob rather than
Alice can carry out unambiguous entanglement concentration, with exactly the
same probability of success.

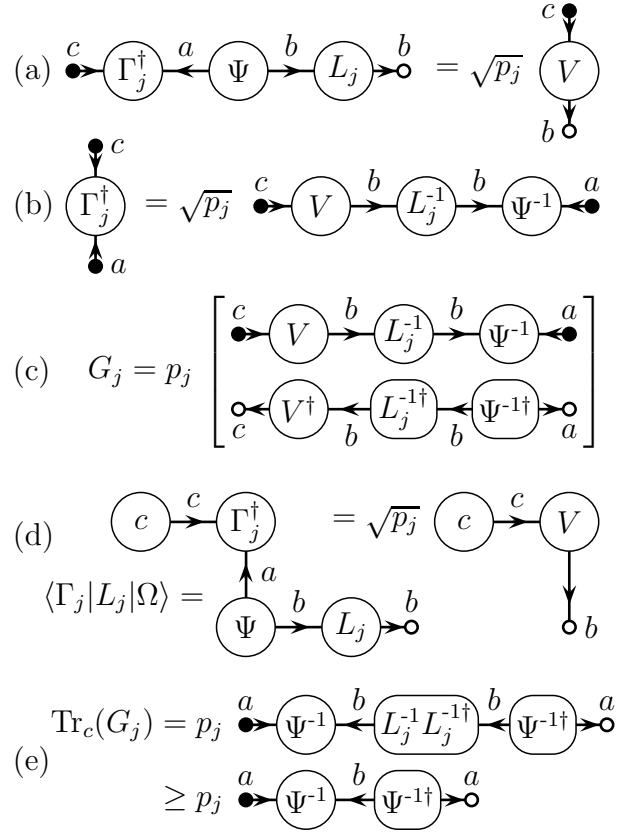
\begin{figure}[h]
$$
\begin{pspicture}(0,-10.1)(8.0,0.9) 
\newpsobject{showgrid}{psgrid}{subgriddiv=1,griddots=10,gridlabels=6pt}
\psset{
labelsep=2.0,
arrowsize=0.150 1,linewidth=\lwd}
\def\mnone{{\hbox{\scriptsize-1}}}
\def\mnonead{{\hbox{\scriptsize-1$\dagger$}}}
\def\dput(#1)#2#3{\rput(#1){#2}\rput(#1){#3}} 
\def\lbrk{\left\{\vvv{0.5}\right.}
\def\lbrkb{\left\{\vvv{0.6}\right.}
\def\obcs{\pscircle[fillcolor=white,fillstyle=solid,linewidth=\lwn](0,0){0.35}}
\def\rbrk{\left.\vvv{0.6}\right\}}
\def\clnr(#1)#2{\dput(#1){\cldot}{\rput[r](-0.2,0){$#2$}}} 
\def\clnl(#1)#2{\dput(#1){\cldot}{\rput[l](0.2,0){$#2$}}} 
\def\clnb(#1)#2{\dput(#1){\cldot}{\rput[b](0,0.2){$#2$}}} 
\def\clnt(#1)#2{\dput(#1){\cldot}{\rput[t](0,-0.2){$#2$}}} 
\def\opnr(#1)#2{\dput(#1){\opdot}{\rput[r](-0.2,0){$#2$}}} 
\def\opnl(#1)#2{\dput(#1){\opdot}{\rput[l](0.2,0){$#2$}}} 
\def\opnb(#1)#2{\dput(#1){\opdot}{\rput[b](0,0.2){$#2$}}} 
\def\opnt(#1)#2{\dput(#1){\opdot}{\rput[t](0,-0.2){$#2$}}} 
\def\rarr(#1){\rput(#1){\psline{->}(-0.2,0)(0,0)}}
\def\larr(#1){\rput(#1){\psline{->}(0.2,0)(0,0)}}
\def\uarr(#1){\rput(#1){\psline{->}(0,-0.2)(0,0)}}
\def\darr(#1){\rput(#1){\psline{->}(0,0.2)(0,0)}}
\def\sbsc{0.9cm}
\def\lfsb{\rput(0,0){$\left[\vrule height \sbsc depth \sbsc width 0pt\right.$}}
\def\rtsb{\rput(0,0){$\left.\vrule height \sbsc depth \sbsc width 0pt\right]$}}
	\def\aobja{
\psline(0,0)(4.4,0)
\dput(0.8,0){\obc}{$\Gm_j\ad$}
\dput(2.2,0){\obc}{$\Psi$}
\dput(3.6,0){\obc}{$L_j$}
\rarr(0.35,0)
\larr(1.35,0)
\rput[b](1.5,0.2){$a$}
\rarr(3.05,0)
\rput[b](2.9,0.2){$b$}
\rarr(4.3,0)
\clnb(0,0){c}
\opnb(4.4,0){b}
	}
	\def\aobjb{
\psline(0,0)(0,1.6)
\dput(0,0.8){\obc}{$V$}
\darr(0,1.25)\darr(0,0.1)
\clnr(0,1.6){c}
\opnr(0,0){b}
	}
	\def\bobja{
\psline(0,0)(0,1.6)
\dput(0,0.8){\obc}{$\Gm_j\ad$}
\darr(0,1.25)\uarr(0,0.35)
\clnl(0,1.6){c}
\clnl(0,0){a}
	}
	\def\bobjb{
\psline(0,0)(4.4,0)
\dput(0.8,0){\obc}{$V$}
\dput(2.2,0){\obc}{$L_j^\mnone$}
\dput(3.6,0){\obc}{$\Psi^\mnone$}
\rarr(0.35,0)
\rarr(1.65,0)
\rput[b](1.5,0.2){$b$}
\rarr(3.05,0)
\rput[b](2.9,0.2){$b$}
\larr(4.05,0)
\clnb(0,0){c}
\clnb(4.4,0){a}
	}
	\def\bobjbad{
\psline(0,0)(4.4,0)
\dput(0.8,0){\obc}{$V\ad$}
\dput(2.2,0){\obr{0.45}}{$L_j^\mnonead$}
\dput(3.55,0){\obr{0.45}}{$\Psi^\mnonead$}
\larr(0.1,0)
\larr(1.35,0)
\rput[t](1.5,-0.2){$b$}
\larr(2.8,0)
\rput[t](2.9,-0.2){$b$}
\rarr(4.3,0)
\opnt(0,0){c}
\opnt(4.4,0){a}
	}
	\def\cobj{
\rput(-0.3,0){\lfsb}
\rput(0,0.5){\bobjb}
\rput(0,-0.5){\bobjbad}
\rput(4.7,0){\rtsb}
	}
	\def\dobja{
\psline(0,1.4)(1.4,1.4)(1.4,0)(2.8,0)(3.6,0)
\dput(0,1.4){\obc}{$c$}
\dput(1.4,1.4){\obc}{$\Gm_j\ad$}
\dput(1.4,0){\obc}{$\Psi$}
\dput(2.8,0){\obc}{$L_j$}
\rarr(0.85,1.4)
\rput[b](0.7,1.6){$c$}
\uarr(1.4,0.85)
\rput[l](1.6,0.7){$a$}
\rarr(2.25,0)
\rput[b](2.1,0.2){$b$}
\rarr(3.5,0)
\opnb(3.6,0){b}
	}
	\def\dobjb{
\psline(0,1.4)(1.4,1.4)(1.4,0)
\dput(0,1.4){\obc}{$c$}
\dput(1.4,1.4){\obc}{$V$}
\rarr(0.85,1.4)
\rput[b](0.7,1.6){$c$}
\darr(1.4,0.3)
\opnl(1.4,0){b}
	}

	\def\eobja{
\psline(0,0)(4.8,0)
\clnb(0,0){a}
\rarr(0.35,0)
\dput(0.8,0){\obc}{$\Psi^\mnone$}
\larr(1.3,0)
\rput[b](1.5,0.2){$b$}
\dput(2.4,0){\obr{0.7}}{$L^\mnone_j L^\mnonead_j$}
\larr(3.2,0)
\rput[b](3.3,0.2){$b$}
\dput(3.95,0){\obr{0.45}}{$\Psi^\mnonead$}
\rarr(4.7,0)
\opnb(4.8,0){a}
	}
	\def\eobjb{
\psline(0,0)(3,0)
\clnb(0,0){a}
\rarr(0.35,0)
\dput(0.8,0){\obc}{$\Psi^\mnone$}
\larr(1.35,0)
\rput[b](1.5,0.2){$b$}
\dput(2.15,0){\obr{0.45}}{$\Psi^\mnonead$}
\rarr(2.9,0)
\opnb(3,0){a}
	}

	\rput(0,0){
\rput[l](0,0){(a)}
\rput(0.8,0){\aobja}
\rput(6.2,0){$=\sqrt{p_j}$}
\rput(7.4,-0.8){\aobjb}
	}

	\rput(0,-1.8){
\rput[l](0,0){(b)}
\rput(1.1,-0.8){\bobja}
\rput(2.33,0){$=\sqrt{p_j}$}
\rput(3.3,0){\bobjb}
	}

	\rput(0,-4){
\rput[l](0,0){(c)}
\rput[l](1.0,0){$G_j = p_j$}
\rput(3.0,0){\cobj}
	}
	
	\rput(0,-6){
\rput[l](0,-0.2){(d)}
\rput[l](0.4,-1){$\lgl\Gm_j|L_j|\Om\rgl =$}
\rput(1.7,-1.4){\dobja}
\rput[r](5.4,0){$=\sqrt{p_j}$}
\rput(6,-1.4){\dobjb}	
	}

	\rput(0,-8.7){
\rput[l](0,-0.5){(e)}
\rput[r](2.8,0){$\Tr_c(G_j)=p_j$}
\rput(3.1,0){\eobja}
\rput[r](2.8,-){$\geq p_j$}
\rput(3.1,-1){\eobjb}
	}

\end{pspicture}
$$
\caption{Diagrams associated with unambiguous teleportation using a POVM
followed by a nonunitary correction.}
\label{fgr9}
\end{figure}

	\subsection{General teleportation protocol}
\label{sct4c}

Consider a protocol in which Alice carries out a POVM $\{G_j\}$, and when Bob
is informed of outcome $j$ he carries out an unambiguous operation $L_j$ with a
probability of success that may be less than 1.  We shall assume that for every
$j\geq 1$, $G_j$ is proportional to a projector onto a pure state.  There is no
harm in making this assumption for a protocol which achieves the maximum
possible probability of success, since any positive operator in a POVM can
always be refined without reducing the amount of information available to
Alice, and thus to Bob.  Hence we write
\begin{equation}
  G_j=|\Gm_j\rgl\lgl\Gm_j| \text{ for } j\geq 1;\quad 
  G_0=I-\sum_{j\geq 1} G_j.
\label{eqn24}
\end{equation}
If Bob's operation $L_j$ is successful, the net result will be a unitary map
$V$ from $\HS_c$ to $\HS_a$ as indicated in Fig.~\ref{fgr9}(a), up to a
constant of proportionality designated by $\sqrt{p_j}$. Note that $V$ must be
independent of $j$ apart from a possible (uninteresting) phase.  Part (b) of
Fig.~\ref{fgr9} is the diagrammatic solution to the equation for $\lgl\Gm_j|$
in (a), and from it one can construct the diagram for $G_j$ in (c).  The
probability that Alice will obtain outcome $j$ \emph{and} Bob will succeed in
carrying out the corresponding operation $L_j$ is given by
\begin{equation}
  \lgl\Om|L_j\ad G_j L_j|\Om\rgl 
  = \lgl\Om|L_j\ad| \Gm_j\rgl\lgl\Gm_j|L_j |\Om\rgl
 =\|\lgl\Gm_j|L_j|\Om\rgl\|^2 = p_j,
\label{eqn25}
\end{equation}
assuming an initial normalized state \eqref{eqn16}.  Here $\lgl\Gm_j|L_j
|\Om\rgl\in\HS_b$ is the ket indicated in Fig.~\ref{fgr9}(d), and the right
side of this figure justifies the final equality in \eqref{eqn25}, since
$|c\rgl$ is normalized and $V$ is unitary.

For the POVM to be physically realizable it must be the case that
\begin{equation}
  \sum_{j\geq 1} G_j \leq I_c\ot I_a,
\label{eqn26}
\end{equation}
which implies that
\begin{equation}
  d\cdot I_a \geq \sum_j  \Tr_c(G_j) \geq
   \Blp\sum_{j\geq 1}p_j\Brp \Psi^{-1}\Psi^{-1\dagger},
\label{eqn27}
\end{equation}
where the second inequality, Fig.~\ref{fgr9}(e), comes about from the fact that
the largest eigenvalue of the positive operator $L_j\ad L_j$ cannot exceed 1
(App.~\ref{sctpa}), so the smallest eigenvalue of its inverse $L^{-1}_j
L^{-1\dagger}_j$ is not less than 1, and therefore $L^{-1}_jL^{-1\dagger}_j\geq
I_b$.  As the largest eigenvalue of the operator $\Psi^{-1}\Psi^{-1\dagger}$ is
$1/\lm_m^2$, \eqref{eqn27} implies that
\begin{equation}
  p_s = \sum_{j\geq 1} p_j \leq d\lm_m^2 = p_c,
\label{eqn28}
\end{equation}
in agreement with the argument given in Sec.~\ref{sct4a}.

	\subsection{Optimum strategies}
\label{sct4d}

Since optimal unambiguous entanglement concentration followed by a standard
teleportation protocol achieves the optimal probability of success, an obvious
way to produce an optimal strategy of the form described in Sec.~\ref{sct4c} is
to produce Alice's POVM by combining the operator $K$ for entanglement
concentration with the fully-entangled basis $\{|\Phi_j\rgl\}$ appropriate for
standard teleportation, setting
\begin{equation}
  |\Gm_j\rgl = (I_c\ot K\ad)|\Phi_j\rgl.
\label{eqn29}
\end{equation}
in \eqref{eqn24}. Using the fact that $\sum_j |\Phi_j\rgl\lgl\Phi_j|$ is the
identity on $\HS_{ac}$, one can check that
\begin{equation}
  \sum_{j\geq 1} G_j = \sum_{j\geq 1}|\Gm_j\rgl\lgl\Gm_j| 
  = I_c\ot K\ad K \leq   I_c\ot I_a,
\label{eqn30}
\end{equation}
as the largest eigenvalue of $K\ad K$ is 1, so the POVM, with $G_0$
given by \eqref{eqn24}, is physically realizable.  Bob, on the other hand,
carries out unitary corrections which always succeed, as in standard
teleportation.  By combining \eqref{eqn29}, \eqref{eqn22} with
$\mu=\sqrt{d}\,\lm_m$, \eqref{eqn8} and \eqref{eqn25}, with $L_j$ and $L_j\ad$
omitted from the last, one finds that $p_j=\lm_m^2/d$, independent of the
initial $|c\rgl$ and independent of $j$, so \eqref{eqn28} is an equality, as
expected.

Obviously, Bob rather than Alice could carry out the entanglement concentration
which takes $|\Psi\rgl$ to $|\Psi_f\rgl$ using an operator $L$ which is the
analog of $K$ in \eqref{eqn22}, and, if successful, utilize the outcome of
Alice's measurements in the fully-entangled basis $\{|\Phi_j\rgl\}$ to apply a
unitary correction $U_j$ in case $j$ to his half of $|\Psi_f\rgl$.  His two
steps can be combined into one by defining an operation
\begin{equation}
  L_j = U_j L.
\label{eqn31}
\end{equation}
Thus in this protocol Alice's POVM is the standard projective measurement in a
fully-entangled basis, whereas Bob's correction $L_j$ will typically not be
unitary, and will sometimes fail.  Because the shared resource is only
partially entangled, the probability that Alice will obtain outcome $j$ depends
in general on the state $|c\rgl$ she is trying to teleport, contrary to what
one might have expected from the general rule (see, for example, \cite{NlCv97})
that teleportation is only possible when neither Alice nor Bob learns anything
about the state being teleported.  However, the \emph{combined} probability
$p_j$ that Alice obtains $j$ \emph{and} Bob succeeds in carrying out $L_j$ is
$\lm_m^2/d$ independent of $|c\rgl$, in agreement with the general rule.  It
is also independent of $j$, so \eqref{eqn28} is again an equality, as
expected.

One can design other optimum procedures by imagining the entanglement
concentration task shared by Alice and Bob; e.g., each could do half of it
based upon
\begin{equation}
  |\Psi^{-1/2}\rgl = \sum_j \lm_j^{-1/2} |\bar a_j\rgl\ot|\bar b_j\rgl.
\label{eqn32}
\end{equation}
After that they carry out measurements and unitary corrections as in the
standard protocol. Upon combining Alice's concentration and measurement steps
in a single POVM, and Bob's concentration and unitary correction steps in
nonunitary operators as in \eqref{eqn31}, one arrives at a combined protocol
with, once again, the optimal probability of success.  There may well be other
possibilities, but it seems clear that in one way or another they will have to
employ the inverse, \eqref{eqn9}, of the partially-entangled ket $|\Psi\rgl$,
and can be understood in terms of how they achieve this.

	\subsection{Literature}
\label{sct4e}

Unambiguous entanglement concentration is our term for the ``Procrustean
method'' introduced by Bennett et al.\ \cite{BBPS96}.  It is often referred to
as distillation or entanglement concentration of a single copy, in contrast to
schemes, again see \cite{BBPS96}, where operations are carried out on multiple
independent systems in the same entangled state.  That the probability of
success in \eqref{eqn23} is the maximum possible using local operations and
classical communication was shown by Vidal \cite{Vdl99} and by Lo and Popescu
\cite{LoPp01}.

The notion of unambiguous teleportation was introduced by Mor in 1996 in
unpublished work \cite{Mr96} for $d=2$, using a protocol in which Alice carries
out a POVM and Bob a unitary correction.  For a complete account including
further developments \cite{MrHr99}, see \cite{BrHM04}.  During this development
the term \emph{conclusive teleportation} acquired a broader sense, and in
\cite{BrHM04} it is ``perfect conclusive teleportation'' that corresponds to
our ``unambiguous teleportation.''  Among other things, these authors obtained
the optimum probability of success, \eqref{eqn28} with $d=2$, and pointed out
the close connection between unambiguous teleportation and entanglement
concentration, as noted above in Sec.~\ref{sct4a}.  Their thinking of a POVM
carried out on an entangled state as producing some sort of a mysterious
choice-at-a-distance (TelePOVM), a perspective derived from Hughston et al.\
\cite{HgJW93}, employs vivid imagery that is not actually necessary for a sober
discussion of teleportation in terms of local quantum mechanics when
conditional probabilities are used in a consistent way; see \cite{Grff02b}, and
Chs.~23 and 24 of \cite{Grff02}.

Nonunitary corrections by Bob in the $d=2$ case, as well as modifications of
Alice's measurement scheme relative to the standard teleportation protocol,
were considered by Li et al.\ \cite{LiLG00} and by Bandyopadhyay
\cite{Bndy00}. Both articles contain explicit protocols which achieve the
optimum probability of success.  By contrast, Agrawal and Pati in
\cite{AgPt02,PtAg04} (the two papers are quite similar) restricted Alice to
projective measurements and Bob to unitary corrections, so their protocol does
not have the optimum probability of success.  However, they made the
interesting observation that the measurement basis states used by Alice must
have the same entanglement as that of the shared resource. From our perspective
this arises from the fact that a $d=2$ bipartite ket and its inverse have the
same entanglement, since they have the same Schmidt coefficients when
normalized in the same way.  However, this is no longer true in higher
dimensions $d\geq3$, where the entanglement of the inverse is in general not
the same as that of the original ket. (The term ``entanglement matching''
occurs in \cite{LiLG00}, but its significance is not clear.)

There have been several studies of unambiguous teleportation in the case of
general $d=d_a=d_b=d_c$, the one we have been considering.  Son et al.\
\cite{SLKP01} and Roa et al.\ \cite{RDFG03} both employ protocols in which
Alice uses a POVM and Bob applies a unitary correction.  In both cases the
probability of success is optimal in the sense that \eqref{eqn28} is an
equality.  However, their arguments that they have optimal procedures are not
easy to follow, and in \cite{RDFG03} the aim seems to be that of maximizing the
\emph{average} fidelity (over all pure input states) of the imperfect
teleportation scheme, rather than maximizing the probability $p_s$ (in our
notation) of conclusive teleportation with unit fidelity.

The approach of Kurucz and colleagues \cite{KrKJ01,KKAJ03} is similar to ours
in focusing attention on the inverse of the partially-entangled resource state.
They construct the inverse by first using the entangled state $|\Psi\rgl$ (in
our notation) to produce an antilinear map from $\HS_a$ to $\HS_b$, which then
has an inverse of the usual sort.  They use this to place conditions on Alice's
measurement, again converted to an antilinear map, so that it will result in a
unitary channel from $\HS_c$ to $\HS_b$, with Bob carrying out a final unitary
correction.  They consider general $d$, but do not address the question of a
measurement strategy that achieves the maximum overall probability of success,
unlike the work cited in the previous paragraph.  The use of antilinear maps,
which were also discussed in connection with teleportation by Uhlmann
\cite{Uhlm03,Uhlm04}, has a certain elegance in that these maps are basis
independent, unlike their linear counterparts, such as the one on the right
side of our Fig.~\ref{fgr5}(d), for which a basis needs to be specified.  In
our own approach the inverse of a bipartite entangled ket is also (when it
exists) a basis-independent concept, so it is not clear to us that antilinear
maps possess a significant conceptual advantage, but we are happy to leave this
to the reader's judgment.

Inverses may also be taken in a manner that makes explicit use of particular
bases, as in the work of Li et al.\ \cite{LiSL02}.  They allow a very general
measurement on Alice's part and also a nonunitary correction by Bob, for a
system of arbitrary dimensionality, as in Sec.~\ref{sct4c} above.  However,
they only consider a single measurement by Alice, and thus do not address the
issue of optimizing unambiguous teleportation.

We have deliberately omitted referring to the substantial literature on some
closely related topics: teleportation using a mixed state as a resource,
teleportation between systems of different dimensionality, and the average
fidelity in cases where the resource is partially entangled.  It may be that
atemporal diagrams are of some use in this broader context, but that remains to
be seen.

	\section{Noisy Quantum Channels}
\label{sct5}

	\subsection{Channel Operators}
\label{sct5a}

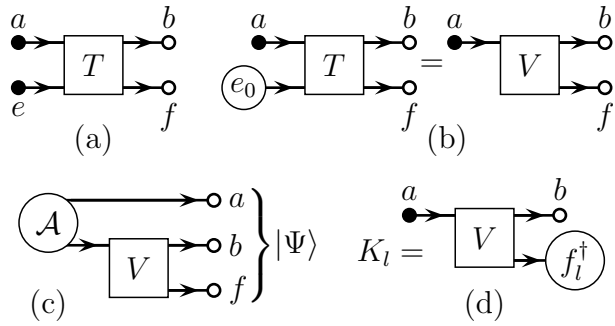
\begin{figure}[h]
$$
\begin{pspicture}(0,-3.2)(8.0,1.2) 
\newpsobject{showgrid}{psgrid}{subgriddiv=1,griddots=10,gridlabels=6pt}
\psset{
labelsep=2.0,
arrowsize=0.150 1,linewidth=\lwd}
\def\lbrk{\left\{\vvv{0.5}\right.}
\def\lbrkb{\left\{\vvv{0.6}\right.}
\def\rbrk{\left.\vvv{0.6}\right\}}
\def\obcs{\pscircle[fillcolor=white,fillstyle=solid,linewidth=\lwn](0,0){0.3}}
	\def\aobj{
\psline(0,0.6)(2,0.6)
\psline{>->}(0.2,0.6)(1.8,0.6)
\psline(0,0)(2,0)
\psline{>->}(0.2,0)(1.8,0)
\rput(1.0,0.3){\obx}\rput(1.0,0.3){$T$}
\rput(0,0.6){\cldot}\rput[b](0,0.8){$a$}
\rput(2,0.6){\opdot}\rput[b](2,0.8){$b$}
\rput(0,0){\cldot}\rput[t](0,-0.2){$e$}
\rput(2,0){\opdot}\rput[t](2,-0.2){$f$}
	}
	\def\bobja{
\psline(0,0.6)(2,0.6)
\psline{>->}(0.2,0.6)(1.8,0.6)
\psline(0,0)(2,0)
\psline{>->}(0.2,0)(1.8,0)
\rput(1.0,0.3){\obx}\rput(1.0,0.3){$T$}
\rput(0,0.6){\cldot}\rput[b](0,0.8){$a$}
\rput(2,0.6){\opdot}\rput[b](2,0.8){$b$}
\rput(-0.2,0){\obcs}\rput(-0.2,0){$e_0$}
\rput(2,0){\opdot}\rput[t](2,-0.2){$f$}
	}
	\def\bobjb{
\psline(0,0.6)(2,0.6)
\psline{>->}(0.2,0.6)(1.8,0.6)
\psline(1,0)(2,0)
\psline{->}(1.2,0)(1.8,0)
\rput(1.0,0.3){\obx}\rput(1.0,0.3){$V$}
\rput(0,0.6){\cldot}\rput[b](0,0.8){$a$}
\rput(2,0.6){\opdot}\rput[b](2,0.8){$b$}
\rput(2,0){\opdot}\rput[t](2,-0.2){$f$}
	}
	\def\cobj{
\psline(1,0)(2.2,0)\psline{->}(1,0)(2.0,0)
\rput(2.2,0){\opdot}\rput[l](2.4,0){$f$}
\psline(0.0,0.6)(2.2,0.6)\psline{>->}(0.4,0.6)(2.0,0.6)
\rput(2.2,0.6){\opdot}\rput[l](2.4,0.6){$b$}
\psline(0.0,1.2)(2.2,1.2)\psline{->}(0.3,1.2)(2.0,1.2)
\rput(2.2,1.2){\opdot}\rput[l](2.4,1.2){$a$}
\rput(1.2,0.3){\obx}\rput(1.2,0.3){$V$}
\rput(0,0.9){\obc}\rput(0,0.9){$\AS$}
\rput(2.8,0.6){$\rbrk$}\rput[l](3.0,0.6){$|\Psi\rgl$}
	}
	\def\dobj{
\psline(0,0.6)(2,0.6)
\psline{>->}(0.2,0.6)(1.8,0.6)
\psline(1,0)(2,0)\psline{->}(1,0)(1.75,0)
\rput(2.2,0){\obc}\rput(2.2,0){$f_l\ad$}
\rput(1.0,0.3){\obx}\rput(1.0,0.3){$V$}
\rput(0,0.6){\cldot}\rput[b](0,0.8){$a$}
\rput(2,0.6){\opdot}\rput[b](2,0.8){$b$}
	}

\rput(0.1,0){\aobj}
\rput[B](1.1,-0.8){(a)}
\rput(3.3,0){\bobja}
\rput(5.6,0.3){$=$}
\rput(5.9,0){\bobjb}
\rput[B](5.8,-0.8){(b)}

\rput(0.5,-2.7){\cobj}
\rput[B](0.5,-3){(c)}
\rput[r](5.5,-2.2){$K_l=$}
\rput(5.3,-2.3){\dobj}
\rput[B](6.3,-3){(d)}

\end{pspicture}
$$
\caption{ Model for a noisy channel based on (a) the unitary map $T$ which
induces the isometry $V$ in (b) if the environment is initially in a pure state
$|e_0\rgl$.  Transposing the isometry yields the channel ket
$|\Psi\rgl$ in (c). A Kraus operator (d) obtained using \eqref{eqn33}.}
\label{fgr10}
\end{figure}

A standard model of a noisy quantum channel is shown in Fig.~\ref{fgr10}.  It
begins with a unitary time-development operator $T$ in (a), which maps the
channel input $\HS_a$ and environment $\HS_e$ to the channel output $\HS_b$ and
environment $\HS_f$ at a later time. Unitarity implies that $\HS_{ae}$ and
$\HS_{bf}$ have equal dimension, but the dimensions $d_a$ and $d_b$ of $\HS_a$
and $\HS_b$ might be different.  (The notation used here follows that of
\cite{Grff05}, with minor modifications, to facilitate comparison.)  The
diagram can be thought of as a quantum circuit to which nodes and arrows have
been added, thus converting it to the corresponding atemporal diagram.  If one
assumes the environment is initially in a pure state $|e_0\rgl$, the
appropriate diagram in part (b) of the figure has three active nodes and
represents the isometry $ V=T|e_0\rgl$ mapping $\HS_a$ to $\HS_{bf}$.  Part (c)
of the figure shows the channel ket $|\Psi\rgl\in\HS_{abf}$ obtained by
applying a transposer $\AS$ to the $a$ node of $V$. (In \cite{Grff05} a
fully-entangled state $|\phi\rgl$ was used instead of $|\AS\rgl$, giving the
same result apart from normalization.) Part (d) is the diagrammatic form for a
Kraus operator
\begin{equation}
 K_l = \lgl f_l|V,
\label{eqn33}
\end{equation}
an element of $\LS_{ba}$, where $\{|f_l\rgl\}$ is some orthonormal basis for
$\HS_f$. (One can, if one wants, think of $K_l$ as associated with a
measurement on $\HS_f$.)

The collection of Kraus operators $\{K_l\}$ depends, of course, on the choice
of basis $\{|f_l\rgl\}$ used to define them. However, the subspace of
$\LS_{ba}$ spanned by their linear combinations is independent of this choice,
and we shall refer to its dimension as the \emph{Kraus rank $\kp$} of the noisy
channel. This subspace is the range of the ``cross operator'' obtained when $V$
is regarded as a map from $\HS_f\ad$ to $\LS_{ba}$, from the bottom to the top
of Fig.~\ref{fgr10}(b), written $V_{ba;f}$ in the notation employed in
Sec.~\ref{sct2e}.  Consequently, the Kraus rank is the (ordinary) rank of this
cross operator,
\begin{equation}
   \kp = \Rn(V_{ba;f}) \leq \min\{ d_ad_b,d_f \},
\label{eqn34}
\end{equation}
where the inequality reflects the fact that the rank of a matrix cannot exceed
the number of rows or the number of columns.

\begin{figure}[h]
$$
\begin{pspicture}(0.25,-5)(17.25,2.5) 
\newpsobject{showgrid}{psgrid}{subgriddiv=1,griddots=10,gridlabels=6pt}
\psset{
labelsep=2.0,
arrowsize=0.150 1,linewidth=\lwd}
\def\dput(#1)#2#3{\rput(#1){#2}\rput(#1){#3}} 
\def\lbrk{\left\{\vvv{0.5}\right.}
\def\lbrkb{\left\{\vvv{0.6}\right.}
\def\rbrk{\left.\vvv{0.6}\right\}}
\def\obcs{\pscircle[fillcolor=white,fillstyle=solid,linewidth=\lwn](0,0){0.35}}
\def\obxb{%
\psframe[fillcolor=white,fillstyle=solid,linewidth=\lwn](-0.5,-0.5)(0.5,0.5)}
\def\clnr(#1)#2{\dput(#1){\cldot}{\rput[r](-0.2,0){$#2$}}} 
\def\clnl(#1)#2{\dput(#1){\cldot}{\rput[l](0.2,0){$#2$}}} 
\def\clnb(#1)#2{\dput(#1){\cldot}{\rput[b](0,0.2){$#2$}}} 
\def\clnt(#1)#2{\dput(#1){\cldot}{\rput[t](0,-0.2){$#2$}}} 
\def\opnr(#1)#2{\dput(#1){\opdot}{\rput[r](-0.2,0){$#2$}}} 
\def\opnl(#1)#2{\dput(#1){\opdot}{\rput[l](0.2,0){$#2$}}} 
\def\opnb(#1)#2{\dput(#1){\opdot}{\rput[b](0,0.2){$#2$}}} 
\def\opnt(#1)#2{\dput(#1){\opdot}{\rput[t](0,-0.2){$#2$}}} 
\def\rarr(#1){\rput(#1){\psline{->}(-0.2,0)(0,0)}}
\def\larr(#1){\rput(#1){\psline{->}(0.2,0)(0,0)}}
\def\uarr(#1){\rput(#1){\psline{->}(0,-0.2)(0,0)}}
\def\darr(#1){\rput(#1){\psline{->}(0,0.2)(0,0)}}
	\def\aobja{
\psline(0,0)(2,0)
\larr(0.2,0)\larr(1.55,0)
\opnt(0,0){a}\clnt(2,0){b}
\psline(0,0.6)(2,0.6)
\rarr(0.42,0.6)\rarr(1.8,0.6)
\clnb(0,0.6){a}\opnb(2,0.6){b}
\dput(1,0.3){\obx}{$Q$}
	}
	\def\aobjb{
\psline(0,0)(2,0)
\larr(0.2,0)\larr(1.55,0)
\opnt(0,0){a}\clnt(2,0){b}
\psline(0,2)(2,2)
\rarr(0.45,2)\rarr(1.8,2)
\clnb(0,2){a}\opnb(2,2){b}
\psarc(0.7,1){0.9}{-40}{40}\psarc{<-}(0.7,1){0.9}{-10}{40}
\rput[l](1.8,1){$f$}
\dput(1.0,0.3){\obx}{$V\ad$}
\dput(1.0,1.7){\obx}{$V$}
	}
	\def\bobja{
\psline(0,0)(2,0)
\larr(0.2,0)\larr(1.55,0)
\opnt(0,0){a}\clnt(2,0){b}
\psline(0,2)(2,2)
\rarr(0.45,2)\rarr(1.8,2)
\clnb(0,2){a}\opnb(2,2){b}
\psline(1,1.5)(2,1.5)\rarr(1.75,1.5)
\dput(2.2,1.5){\obcs}{$f_l\ad$}
\psline(1,0.5)(2,0.5)\larr(1.45,0.5)
\dput(2.2,0.5){\obcs}{$f_l$}
\dput(1.0,0.3){\obx}{$V\ad$}
\dput(1.0,1.7){\obx}{$V$}
\rput(1,1){$\ot$}
	}
	\def\bobjb{
\psline(0,0.3)(2,0.3)
\larr(0.2,0.3)\larr(1.55,0.3)
\opnt(0,0.3){a}\clnt(2,0.3){b}
\psline(0,1.7)(2,1.7)
\rarr(0.45,1.7)\rarr(1.8,1.7)
\clnb(0,1.7){a}\opnb(2,1.7){b}
\dput(1,0.3){\obc}{$K_l\ad$}
\dput(1,1.7){\obc}{$K_l$}

\rput(1,1){$\ot$}
	}
	\def\cobja{
\psline(0,0)(0,2)
\uarr(0,0.45)\uarr(0,1.8)
\dput(0,1){\obc}{$A$}
\psline(0,0)(0.6,0.7)(2.4,0.7)
\psline(0,2)(0.6,1.3)(2.4,1.3)
\larr(0.75,0.7)\rarr(1.0,1.3)
\larr(2.0,0.7)\rarr(2.25,1.3)
\clnr(0,0){a}\opnr(0,2){a}
\opnt(0.6,0.7){a}\clnb(0.6,1.3){a}
\clnt(2.4,0.7){b}\opnb(2.4,1.3){b}
\dput(1.5,1){\obx}{$Q$}
	}
	\def\cobjb{
\psline(0,0)(0,2)
\uarr(0,0.45)\uarr(0,1.8)
\dput(0,1){\obc}{$A$}
\psline(0,0)(0.6,0.3)(2.4,0.3)
\psline(0,2)(0.6,1.7)(2.4,1.7)
\larr(0.75,0.3)\rarr(1.0,1.7)
\larr(2.0,0.3)\rarr(2.25,1.7)
\dput(1.5,0.3){\obc}{$K_l\ad$}
\dput(1.5,1.7){\obc}{$K_l$}
\clnr(0,0){a}\opnr(0,2){a}
\opnt(0.6,0.3){a}\clnb(0.6,1.7){a}
\clnt(2.4,0.3){b}\opnb(2.4,1.7){b}
	}
	\def\dobja{
\psline(0,0)(2,0)
\rarr(0.42,0)\rarr(1.8,0)
\clnt(0,0){b}\opnt(2,0){b}
\psline(0,0.6)(2,0.6)
\rarr(0.42,0.6)\rarr(1.8,0.6)
\clnb(0,0.6){a}\opnb(2,0.6){a}
\dput(1,0.3){\obx}{$R$}
	}
	\def\dobjb{
\psline(-0.4,-1)(-0.4,1.4)(-1.3,1.4)
\rarr(-0.9,1.4)\uarr(-0.4,0.9)\uarr(-0.4,-0.6)
\clnt(-1.3,1.4){a}\clnr(-0.4,-1){b}
\psline(0.4,-1)(0.4,1.4)(1.3,1.4)
\rarr(1.1,1.4)\darr(0.4,0.65)\darr(0.4,-0.85)
\opnt(1.3,1.4){a}\opnl(0.4,-1){b}
\dput(0,0){\obxb}{$Q$}
\dput(-0.4,1.4){\obcs}{$\AS\ad$}
\dput(0.4,1.4){\obcs}{$\AS$}
\rput[l](-0.2,0.75){$a$}\rput[l](0.6,0.75){$a$}
	}
	\def\dobjc{
\psline(-0.9,-1)(-0.9,1.4)(-1.8,1.4)
\rarr(-1.4,1.4)\uarr(-0.9,0.85)\uarr(-0.9,-0.55)
\psline(0.9,-1)(0.9,1.4)(1.8,1.4)
\rarr(1.6,1.4)\darr(0.9,0.6)\darr(0.9,-0.8)
\psarc(0,0.3){0.9}{-140}{-40}\psarc{<-}(0,0.3){0.9}{-100}{-40}
\dput(-0.9,1.4){\obcs}{$\AS\ad$}
\dput(-0.6,0){\obx}{$V\ad$}
\dput(0.9,1.4){\obcs}{$\AS$}
\dput(0.6,0){\obx}{$V$}
\clnt(-1.8,1.4){a}\clnr(-0.9,-1){b}
\opnt(1.8,1.4){a}\opnr(0.9,-1){b}
\rput[t](0,-0.8){$f$}
\rput[l](-0.7,0.75){$a$}\rput[l](1.1,0.75){$a$}
	}

	\rput(2,0){
\rput(0,0.7){\aobja}
\rput(2.5,1){$=$}
\rput(3,0){\aobjb}
\rput[B](2.5,-0.8){(a)}

\rput[l](6,1){$=\sum_l$}
\rput(7.1,0){\bobja}
\rput[l](10,1){$=\sum_l$}
\rput(11.1,0){\bobjb}
\rput[B](10,-0.8){(b)}
	}		

	\rput(0.4,-4){
\rput(0,0){\cobja}
\rput(3.2,1){$=\sum_l$}
\rput(4.3,0){\cobjb}
\rput[B](3.3,-0.8){(c)}
	}

	\rput(-0.3,0){
\rput(8.3,-3.5){\dobja}
\rput(10.8,-3.2){$=$}
\rput(11.8,-3.2){\dobjb}
\rput(13.4,-3.2){$=$}
\rput(15.5,-3.2){\dobjc}
\rput[B](10.6,-4.8){(d)}
	}

\end{pspicture}
$$
\caption{ (a) Definition of the transition operator $Q$. (b) Equivalent
expressions for $Q$ in terms of Kraus operators. (c) Contraction with $Q$
yields the superoperator map $\VS(A)$, see Eq.~\eqref{eqn35}. (d) The dynamical
operator $R$ is the partial transpose of $Q$}
\label{fgr11}
\end{figure}
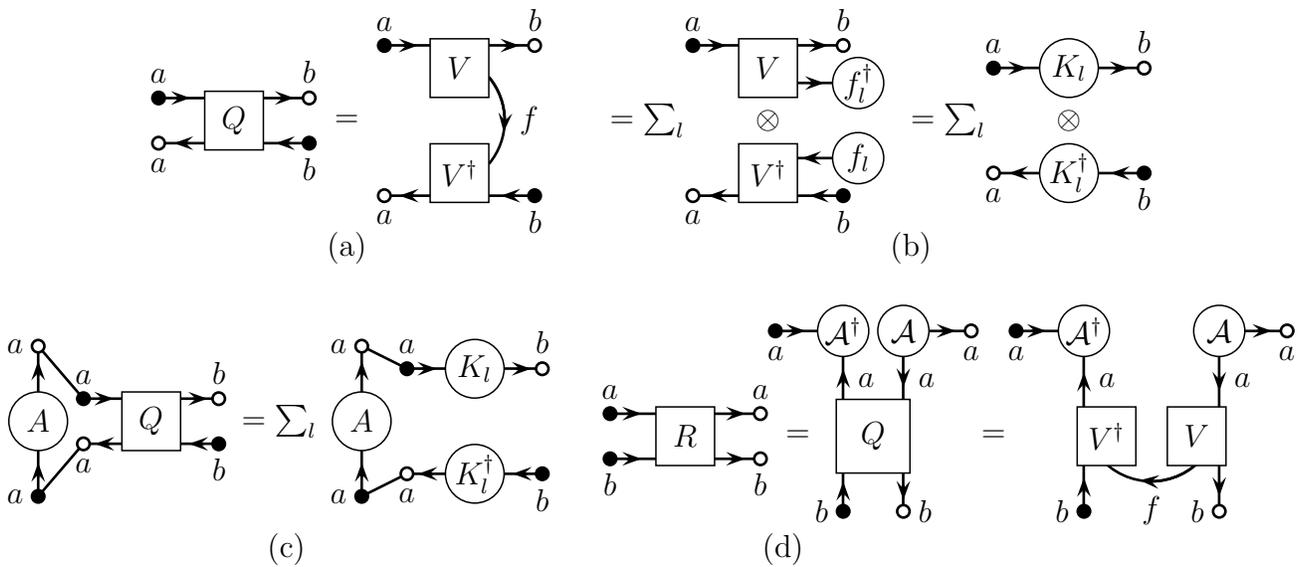

A channel is usually discussed in terms of the superoperator $\VS$ that maps
$\hat\HS_a$ to $\hat\HS_b$, in particular it maps density operators at the
channel entrance to density operators at the channel output.  Two ways of
writing it are:
\begin{equation}
  \VS(A) = \Tr_a[(A\ot I)Q] = \sum_l K_l A K\ad_l,
\label{eqn35}
\end{equation}
where $Q\in\hat\HS_a\ot\hat\HS_b$ is the \emph{transition operator}, in the
notation of \cite{Grff05}, and the Kraus operators were defined in
\eqref{eqn33} above. The object $Q$ corresponds to an atemporal diagram with
four legs obtained by contracting the tensor product of $V$ with its adjoint
$V\ad$, Fig.~\ref{fgr11}(a). By replacing the $f$ contraction line, thought of
as $I_f$, with $\sum_l |f_l\rgl\lgl f_l|$, see Fig.~\ref{fgr3}(e), one obtains
$Q$ in the form shown in Fig.~\ref{fgr11}(b) as a sum of tensor products of
Kraus operators. Thus parts (a) and (b) of the figure correspond to writing $Q$
as
\begin{equation}
 Q=\Tr_f(V\ot V\ad) = \sum_l K_l\ot K_l\ad.
\label{eqn36}
\end{equation}
The meaning of these equations is probably clearest if they are used in
conjunction with the atemporal diagrams in part (a) and (b) of
Fig.~\ref{fgr11}. The way in which $Q$ generates the superoperator $\VS$ in
\eqref{eqn35}, as a map from the left pair of nodes to the right pair of nodes,
is indicated in part (c) of this figure; note how contraction with the operator
$A$ yields a diagram with the open and closed $b$ nodes signifying an operator
on $\HS_b$.

The \emph{dynamical operator}
\begin{equation}
  R=\AS\ad Q\AS
\label{eqn37}
\end{equation}
is the partial transpose of $Q$ in the basis $\{|a_j\rgl\}$, \eqref{eqn5}, and
a map of $\HS_{ab}$ onto itself. The corresponding diagram is shown in
Fig.~\ref{fgr11}(d), where $R$ acts from left to right, and $Q$ has been
rotated by $90\dg$ relative to its orientation in (a).  The left-to-right
reflection symmetry of the final diagram shows that $R$ is a positive
(semidefinite) operator, of the general form $WW\ad$, where $W=V\AS$, regarded
as a map from $\HS_f\ad$ to $\HS_{ab}$, may be thought of as a transposed form
of the $V_{ba;f}$ cross operator introduced previously. Thus the rank of $W$ is
equal to the Kraus rank $\kp$, and since the rank of $WW\ad$ is the same as
that of $W$ (see, e.g., p.~13 of \cite{HrJh85}), the Kraus rank can also be
defined, as in \cite{Grff05}, to be the (ordinary) rank of the dynamical
operator $R$. This in turn is the rank of $Q$ when it is regarded as a map
from $\LS_{ba}$ to $\LS_{ba}$, i.e., bottom-to-top in Fig.~\ref{fgr11}(a).

	\subsection{Complete positivity}
\label{sct5b}

\begin{figure}[h]
$$
\begin{pspicture}(0,-10)(8.0,2.3) 
\newpsobject{showgrid}{psgrid}{subgriddiv=1,griddots=10,gridlabels=6pt}
\psset{
labelsep=2.0,
arrowsize=0.150 1,linewidth=\lwd}
\def\dput(#1)#2#3{\rput(#1){#2}\rput(#1){#3}} 
\def\lbrk{\left\{\vvv{0.5}\right.}
\def\lbrkb{\left\{\vvv{0.6}\right.}
\def\obcs{\pscircle[fillcolor=white,fillstyle=solid,linewidth=\lwn](0,0){0.35}}
\def\rbrk{\left.\vvv{0.6}\right\}}
\def\clnr(#1)#2{\dput(#1){\cldot}{\rput[r](-0.2,0){$#2$}}} 
\def\clnl(#1)#2{\dput(#1){\cldot}{\rput[l](0.2,0){$#2$}}} 
\def\clnb(#1)#2{\dput(#1){\cldot}{\rput[b](0,0.2){$#2$}}} 
\def\clnt(#1)#2{\dput(#1){\cldot}{\rput[t](0,-0.2){$#2$}}} 
\def\opnr(#1)#2{\dput(#1){\opdot}{\rput[r](-0.2,0){$#2$}}} 
\def\opnl(#1)#2{\dput(#1){\opdot}{\rput[l](0.2,0){$#2$}}} 
\def\opnb(#1)#2{\dput(#1){\opdot}{\rput[b](0,0.2){$#2$}}} 
\def\opnt(#1)#2{\dput(#1){\opdot}{\rput[t](0,-0.2){$#2$}}} 
\def\rarr(#1){\rput(#1){\psline{->}(-0.2,0)(0,0)}}
\def\larr(#1){\rput(#1){\psline{->}(0.2,0)(0,0)}}
\def\uarr(#1){\rput(#1){\psline{->}(0,-0.2)(0,0)}}
\def\darr(#1){\rput(#1){\psline{->}(0,0.2)(0,0)}}
	\def\aobj{
\psline(1,0)(2,0)
\psline{-<}(1.2,0)(1.8,0)
\psline(1,2)(2,2)
\psline{->}(1.2,2)(1.8,2)
\psarc(0.7,1){0.9}{-40}{40}\psarc{<-}(0.7,1){0.9}{-10}{40}
\rput[l](1.8,1){$f$}
\psline(-0.5,0.5)(-0.5,1.5)
\psarc(0.5,1.0){1.0}{-160}{-40}\psarc{<-}(0.5,1.0){1.0}{-120}{-40}
\psarc(0.5,1.0){1.0}{40}{160}\psarc{<-}(0.5,1.0){1.0}{100}{160}
\dput(-0.5,0.5){\obc}{$N$}
\dput(-0.5,1.5){\obc}{$N\ad$}
\dput(1.0,0.3){\obx}{$V\ad$}
\dput(1.0,1.7){\obx}{$V$}
\opnl(2,2){b}
\clnl(2,0){b}
\rput[b](0.2,0.2){$a$}
\rput[t](0.2,1.8){$a$}
	}
	\def\bobj{
\psline(0,0)(2,0)
\psline{<-<}(0.2,0)(1.8,0)
\psline(0,2)(2,2)
\psline{>->}(0.2,2)(1.8,2)
\psarc(0.7,1){0.9}{-40}{40}\psarc{<-}(0.7,1){0.9}{-10}{40}
\rput[l](1.8,1){$f$}
\dput(1.0,0.3){\obx}{$V\ad$}
\dput(1.0,1.7){\obx}{$V$}
\rput(0,0){\opdot}\rput[r](-0.2,-0.0){$a$}
\rput(0,2){\cldot}\rput[r](-0.2,2.0){$a$}
\rput(2,0){\cldot}\rput[l](2.2,-0.0){$b$}
\rput(2,2){\opdot}\rput[l](2.2,2.0){$b$}
\psline(0,-0.5)(2,-0.5)
\psline{<-}(0.9,-0.5)(2,-0.5)
\psline(0,2.5)(2,2.5)
\psline{->}(0.0,2.5)(1.1,2.5)
\rput(0,-0.5){\opdot}\rput[r](-0.2,-0.5){$s$}
\rput(2,-0.5){\cldot}\rput[l](2.2,-0.5){$s$}
\rput(0,2.5){\cldot}\rput[r](-0.2,2.5){$s$}
\rput(2,2.5){\opdot}\rput[l](2.2,2.5){$s$}
	}
	\def\cobj{
\psline(1,0)(2,0)
\psline{-<}(1.2,0)(1.8,0)
\psline(1,2)(2,2)
\psline{->}(1.2,2)(1.8,2)
\psarc(0.7,1){0.9}{-40}{40}\psarc{<-}(0.7,1){0.9}{-10}{40}
\rput[l](1.8,1){$f$}
\psline(-0.5,0.5)(-0.5,1.5)
\psarc(0.7,1.0){1.0}{-160}{-90}
\psarc{<-}(0.7,1.0){1.0}{-130}{-90}
\psarc(0.7,1.0){1.0}{90}{160}
\psarc{<-}(0.7,1.0){1.0}{110}{160}
\psarc(0.7,1.0){1.5}{-160}{-90}
\psarc{<-}(0.7,1.0){1.5}{-130}{-90}
\psarc(0.7,1.0){1.5}{90}{160}
\psarc{<-}(0.7,1.0){1.5}{110}{160}
\rput[b](0.2,0.3){$a$}
\rput[t](0.2,1.7){$a$}
\dput(-0.5,0.5){\obc}{$X$}
\dput(-0.5,1.5){\obc}{$X\ad$}
\dput(1.0,0.3){\obx}{$V\ad$}
\dput(1.0,1.7){\obx}{$V$}
\rput(2,0){\cldot}\rput[l](2.2,-0.0){$b$}
\rput(2,2){\opdot}\rput[l](2.2,2.0){$b$}
\psline(0.7,-0.5)(2,-0.5)
\psline(0.7,2.5)(2,2.5)
\rput(2,-0.5){\cldot}\rput[l](2.2,-0.5){$s$}
\rput(2,2.5){\opdot}\rput[l](2.2,2.5){$s$}
\rput[t](-0.3,-0.3){$s$}
\rput[b](-0.3,2.3){$s$}
	}
	\def\dobja{
\psline(0,1.2)(2,1.2)
\psline(1,0.3)(2,0.3)
\psline(1,-0.3)(2,-0.3)
\psline(0,-1.2)(2,-1.2)
\psarc(0.6,0.9){0.6}{-180}{-85}
\psarc{->}(0.6,0.9){0.6}{-180}{-100}
\psarc(0.6,-0.9){0.6}{85}{180}
\psarc{->}(0.6,-0.9){0.6}{85}{130}
\dput(1,0){\obx}{$Q$}
\dput(0,0.9){\obc}{$\AS$}
\dput(0,-0.9){\obc}{$\AS\ad$}
\rput(2,1.2){\opdot}\rput[l](2.2,1.2){$a$}
\rput(2,0.3){\opdot}\rput[l](2.2,0.3){$b$}
\rput(2,-0.3){\cldot}\rput[l](2.2,-0.3){$b$}
\rput(2,-1.2){\cldot}\rput[l](2.2,-1.2){$a$}
\rarr(1.1,1.2)
\rarr(1.85,0.3)
\larr(1.6,-0.3)
\larr(0.9,-1.2)
	}
	\def\dobjb{
\psline(0,-0.6)(0,0.6)
\psline(-0.9,0.6)(0.9,0.6)
\psline(-0.9,-0.6)(0.9,-0.6)
\dput(0,0.6){\obc}{$Y\ad$}
\dput(0,-0.6){\obc}{$Y$}
\clnr(-0.9,-0.6){a}
\clnl(0.9,-0.6){b}
\opnr(-0.9,0.6){a}
\opnl(0.9,0.6){b}
\larr(-0.75,0.6)
\rarr(0.75,0.6)
\rarr(-0.5,-0.6)
\larr(0.5,-0.6)
	}
	\def\eobja{
\psline(0,0.3)(2,0.3)
\psline(0,-0.3)(2,-0.3)
\dput(1,0){\obx}{$Q$}
\rput(2,0.3){\opdot}\rput[l](2.2,0.3){$b$}
\rput(2,-0.3){\cldot}\rput[l](2.2,-0.3){$b$}
\rput(0,0.3){\cldot}\rput[r](-0.2,0.3){$a$}
\rput(0,-0.3){\opdot}\rput[r](-0.2,-0.3){$a$}
\larr(0.2,-0.3)
\rarr(0.5,0.3)
\larr(1.5,-0.3)
\rarr(1.8,0.3)
	}
	\def\eobjb{
\psline(0,-0.6)(0,0.6)
\psline(-1.9,0.6)(0.9,0.6)
\psline(-1.9,-0.6)(0.9,-0.6)
\dput(0,0.6){\obc}{$Y\ad$}
\dput(0,-0.6){\obc}{$Y$}
\clnl(0.9,-0.6){b}
\opnl(0.9,0.6){b}
\rarr(0.75,0.6)
\larr(0.5,-0.6)
\clnr(-1.9,0.6){a}
\opnr(-1.9,-0.6){a}
\rarr(-1.5,0.6)
\larr(-1.75,-0.6)
\dput(-1,0.6){\obc}{$\AS\ad$}
\dput(-1,-0.6){\obc}{$\AS$}
	}

\rput[l](0,1){(a)}
\rput(3.0,0){\aobj}

\rput[l](0,-2.5){(b)}
\rput(1,-3.5){\bobj}

\rput[l](4,-2.5){(c)}
\rput(5.5,-3.5){\cobj}

\rput[l](0,-6){(d)}
	\rput(0.6,0){
\rput(1,-6){\dobja}
\rput(3.8,-6){$=$}
\rput(5.6,-6){\dobjb}
	}

\rput[l](0,-9){(e)}
\rput(1.3,-9){\eobja}
\rput(4.1,-9){$=$}
\rput(6.7,-9){\eobjb}
\end{pspicture}
$$
\caption{Argument that complete positivity of the superoperator $\VS$
corresponding to the transition operator $Q$ is equivalent to positivity of the
``cross operator'' mapping $\LS_{ba}$ to itself.}
\label{fgr12}
\end{figure}
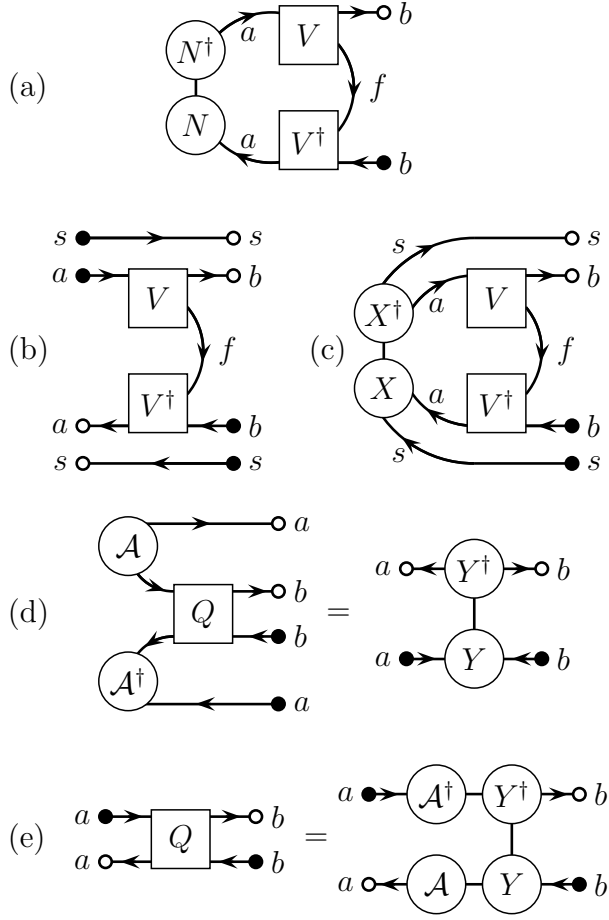

As is well known,  the superoperator $\VS$ for a quantum channel is a
\emph{completely positive} map from $\hat\HS_a$ to $\hat\HS_b$ in the following
sense. Let $\hat\HS_s$ be the space of operators on a Hilbert space $\HS_s$
distinct from any of those we have been considering, and define the
superoperator $\WS$ mapping $\hat\HS=\hat\HS_a\ot\hat\HS_s$ to
$\hat\HS'=\hat\HS_b\ot\hat\HS_s$ through
\begin{equation}
 \WS(A\ot S) = \VS(A)\ot S,
\label{eqn38}
\end{equation}
for any $S$ in $\hat\HS_s$; i.e., $\WS$ is the tensor product of $\VS$ with an
identity map on $\hat\HS_s$.  Then $\VS$ is defined to be a \emph{completely
positive map} provided any superoperator $\WS$ constructed in this manner is a
\emph{positive map}, in the sense that whenever $P\in\hat\HS$ is a positive
(semidefinite) operator, $\WS(P)\in\hat\HS'$ is also a positive (semidefinite)
operator.

This definition is neither simple nor constructive: it provides no direct way
to check whether some superoperator is or is not completely positive.  The
diagram in Fig.~\ref{fgr11}(a) suggests a simpler characterization. If one
thinks of $Q$ as acting from ``bottom to top,'' which means as a map of
$\LS_{ba}$ to itself, the atemporal diagram has the characteristic symmetry of
a positive (definite) operator, see Sec.~\ref{sct2d} and the examples in
Fig.~\ref{fgr4}(b) and (e).  In fact, the positivity of this ``cross operator''
of the transition operator $Q$ is a necessary and sufficient condition that
$\VS$ be a completely positive map, as first pointed out by Choi. (See Thm.~2
of \cite{Choi75}; the reader who finds the argument difficult to follow may
wish to consult \cite{Lng03}.)  A diagrammatic proof of this result is given in
Fig.~\ref{fgr12}.

Figure~\ref{fgr12}(a) shows that if $A$ is a positive operator on $\HS_a$,
$\VS(A)$ is a positive operator on $\HS_b$.  Here $A$ has been written in the
form $N\ad N$, see Fig.~\ref{fgr4}(b), and one can see that the operator on
$\HS_b$ which results from contraction with $Q$ is positive, because of the
(top-to-bottom) reflection symmetry of the diagram.  The transition operator
corresponding to the superoperator $\WS$ in \eqref{eqn38} is shown in
Fig.~\ref{fgr12}(b), where the two horizontal lines representing the identity
superoperator on $\hat\HS_s$ are drawn in a way which preserves the
top-to-bottom reflection symmetry.  Thus when $\WS$ is applied to an arbitrary
positive operator, shown in (c) as $X\ad X$, the result is an operator which is
positive.  Note that any positive operator on $\LS_{ba}$ can be expressed in
the form of a diagram similar to that used for $Q$ in Fig.~\ref{fgr11}(a), with
$V$ some object that need not be an isometry. Thus what we have shown is that
if the ``cross operator'' of the transition operator defining any superoperator
$\VS$ is positive (semidefinite), the superoperator $\VS$ itself is completely
positive, according to the standard definition based on \eqref{eqn38}.

The converse argument, from $\WS$ in the form \eqref{eqn38} as a positive map
to the positivity of $Q$ as an operator on $\LS_{ba}$, begins in
Fig.~\ref{fgr12}(d).  We have chosen $\HS_s$ to be a copy of $\HS_a$, and
allowed $\WS$ to act on a particular operator $\AS\ot\AS\ad$ formed from the
transposer defined in Sec.~\ref{sct2e} and its adjoint, which by symmetry is a
positive (semidefinite) operator on $\HS_a\ot\HS_a$.  Since by assumption $\WS$
maps positive operators to positive operators, the result must be a positive
operator on $\HS_{ab}$, and thus of the form $Y\ad Y$ shown on the right side
of (d). The equation for $Q$ in (d) is solved in (e), and the symmetry of the
diagram on the right side (read from bottom to top) implies that the cross
operator of $Q$ acting on $\LS_{ba}$ is positive semidefinite. This completes
the proof of equivalence.

 In constructing the proof we have, incidentally, proven that the complete
positivity of $\VS$ is equivalent to the positivity of the \emph{dynamical
operator} $R=\AS\ad Q\AS$, the partial transpose of the transition operator $Q$
with respect to the orthonormal basis of $\HS_a$ giving rise to $\AS$.
While the preceding results are not new, the diagrammatic analysis makes the
derivation particularly simple by showing that everything hinges on properties
of a single ``object'' $Q$  viewed in various different ways.

	\subsection{Channel based on mixed-state environment}
\label{sct5c}

For simplicity the following discussion is limited to the case of a noisy
channel in which the input and output have the same dimension $d_b=d_a$.  In
order to model this channel using a unitary $T$ applied to the system together
with an environment initially in a pure state, as in Fig.~\ref{fgr10}(b), the
dimension $d_e=d_f$ of the environment must be at least as great as the Kraus
rank $\kp$; see the remarks in Sec.~\ref{sct5a}.  But $\kp$ cannot exceed
$d_ad_b=d_a^2$, consistent with the well-known fact that using an environment
of dimension $d_e=d_a^2$ and a suitable $T$ one can model any channel of this
sort.

However, if the environment is initially in a mixed state $\rho_e$, $d_e^2$
rather than $d_e$ becomes the upper bound on $\kp$ (see below following
\eqref{eqn40}), and thus one might have supposed that any noisy channel could
be modeled using an environment of dimension $d_e=d_a$, rather than
$d_e=d_a^2$, by allowing it to be initially in a mixed state.  In fact this is
not the case \cite{ZlRf02}, but good criteria for distinguishing channels which
can and cannot be modeled using an environment of dimension $d_e < d_a^2$ in a
mixed state are not known at present.  In the case $d_e=d_a=2$, numerical
evidence \cite{Trao99} indicates that $\kp$ for such a channel can only take on
the values 1, 2, and 4; 3 is excluded.  We shall present a simple proof of this
result based on a diagrammatic analysis which, while not resolving the general
mixed-environment channel problem, does focus attention on what could be a
helpful tool: properties of the cross operator corresponding to a unitary
operator on a bipartite system.

\begin{figure}[h]
$$
\begin{pspicture}(0,-0.7)(8.0,3.2) 
\newpsobject{showgrid}{psgrid}{subgriddiv=1,griddots=10,gridlabels=6pt}
\psset{
labelsep=2.0,
arrowsize=0.150 1,linewidth=\lwd}
\def\dput(#1)#2#3{\rput(#1){#2}\rput(#1){#3}} 
\def\obcs{\pscircle[fillcolor=white,fillstyle=solid,linewidth=\lwn](0,0){0.35}}
\def\lfsb{\rput(0,0){$\left[\vrule height 1.0cm depth 1.0cm width 0pt\right.$}}
\def\rtsb{\rput(0,0){$\left.\vrule height 1.0cm depth 1.0cm width 0pt\right]$}}

\def\clnr(#1)#2{\dput(#1){\cldot}{\rput[r](-0.2,0){$#2$}}} 
\def\clnl(#1)#2{\dput(#1){\cldot}{\rput[l](0.2,0){$#2$}}} 
\def\clnb(#1)#2{\dput(#1){\cldot}{\rput[b](0,0.2){$#2$}}} 
\def\clnt(#1)#2{\dput(#1){\cldot}{\rput[t](0,-0.2){$#2$}}} 
\def\opnr(#1)#2{\dput(#1){\opdot}{\rput[r](-0.2,0){$#2$}}} 
\def\opnl(#1)#2{\dput(#1){\opdot}{\rput[l](0.2,0){$#2$}}} 
\def\opnb(#1)#2{\dput(#1){\opdot}{\rput[b](0,0.2){$#2$}}} 
\def\opnt(#1)#2{\dput(#1){\opdot}{\rput[t](0,-0.2){$#2$}}} 
\def\rarr(#1){\rput(#1){\psline{->}(-0.2,0)(0,0)}}
\def\larr(#1){\rput(#1){\psline{->}(0.2,0)(0,0)}}
\def\uarr(#1){\rput(#1){\psline{->}(0,-0.2)(0,0)}}
\def\darr(#1){\rput(#1){\psline{->}(0,0.2)(0,0)}}
	\def\aobj{
\psline(0,0.5)(0,2.5)
\psline(0,2.5)(2,2.5)
\psline(0,3.1)(2,3.1)
\psline(0.5,0)(2,0)
\psarc(0.5,0.5){0.5}{-180}{-90}
\uarr(0,1.8)\rarr(0.45,2.5)\rarr(0.45,3.1)
\rarr(1.8,2.5)\rarr(1.8,3.1)
\rarr(1.15,0)
\dput(0,1){\obc}{$\Phi$}
\dput(1,2.8){\obx}{$U$}
\opnr(0,2){e} \clnr(0,2.5){e} \clnr(0,3.1){a}
\opnl(2,0){g} \opnl(2,2.5){c} \opnl(2,3.1){b}
	}
	\def\bobj{
\psline(0,0)(0,2)
\psline(2,0)(2,2)
\psline(0,2.5)(2,2.5)
\psline(0,3.1)(2,3.1)
\darr(0,0.2)\uarr(0,1.8)\rarr(0.45,2.5)\rarr(0.45,3.1)
\darr(2,0.85)\rarr(1.8,2.5)\rarr(1.8,3.1)
\psline[linestyle=dashed](0,2)(0,2.5)
\psline[linestyle=dashed](2,2)(2,2.5)
\dput(0,1){\obc}{$\Phi$}
\dput(1,2.8){\obx}{$U$}
\opnr(0,0){g} \opnr(0,2){e} \clnr(0,2.5){e} \clnr(0,3.1){a}
\opnl(2,0){c} \clnl(2,2){c} \opnl(2,2.5){c} \opnl(2,3.1){b}
\rput(1,1){$\bld{\ot}$}
\rput(-0.7,1){\lfsb}\rput(2.7,1){\rtsb}
\rput[r](-0.9,1){$S=$}
	}


\rput(0.5,0){\aobj}
\rput[B](1.5,-0.6){(a)}
\rput(5.2,0){\bobj}
\rput[B](6.2,-0.6){(b)}

\end{pspicture}
$$
\caption{Channel based on mixed-state environment obtained by tracing
$|\Phi\rgl\lgl\Phi|$ over $\HS_g$. In (b) the atemporal diagram in (a)
is redrawn in a way which helps interpreting it as a cross operator.}
\label{fgr13}
\end{figure}
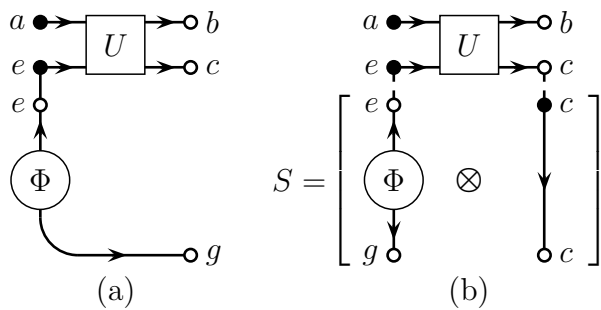

Figure~\ref{fgr13}(a) shows the situation we wish to consider as an atemporal
diagram drawn in a way which makes it resemble a quantum circuit.  The initial
mixed state $\rho_e$ of the environment has been ``purified'' in the usual
fashion by introducing a hypothetical reference system $\HS_g$ with dimension
$d_g=d_e$, so that $\rho_e$ is the partial trace over $\HS_g$ of an entangled
state $|\Phi\rgl$ on $\HS_{eg}$.  Thus we are back to the case considered in
Fig.~\ref{fgr10}: an (enlarged) environment in an initial pure state, with a
time development operator $T=U\ot I_g$, where $U$ is a unitary operator acting
on $\HS_{ae}$, while the final environment Hilbert space $\HS_f$ in that figure
is now $\HS_{cg}$, with $d_c=d_e$.

The Kraus rank $\kp$ of this channel is the rank of the cross operator
$V_{ba;f}$, \eqref{eqn34}, and in studying what values it can take it is
helpful to represent it in the form, see Fig.~\ref{fgr13}(b),
\begin{equation}
 V_{ba;f}=  V_{ba;cg} =  U_{ba;ce} S_{ce;cg} 
\label{eqn39}
\end{equation}
of a product of the cross operator (bottom-to-top in the figure) $U_{ba;ce}$
for the unitary $U$ and a second (bottom-to-top) map $S_{ce;cg}$ corresponding
to the object $S$ defined in the figure. In fact, $S_{ce;cg}$ is a tensor
product of the map generated by $|\Phi\rgl$, times the identity on $\HS_c$, and
thus its rank is $d_c=d_e$ times the Schmidt rank $\sg$ of $|\Phi\rgl$.  Since
the rank of the matrix product of two matrices cannot exceed that of either
one, \eqref{eqn39} leads to the conclusion that
\begin{equation}
  \kp \leq \min\{\Rn(U_{ba;ce}),d_e\sg\}
\label{eqn40}
\end{equation}
in the notation of Sec.~\ref{sct2e}, where $\Rn()$ stands for rank.  If
$\sg=1$, meaning $|\Phi\rgl$ is a product state and the environment
$\HS_e$ is initially in a pure state, \eqref{eqn40} tells us that $\kp$ cannot
be larger than $d_e$.  The other extreme is $\sg=d_e=d_g$, so that
$d_e^2$ replaces $d_e$ as a bound, and since $S_{ce;cg}$ is in this case
nonsingular, $\kp=\Rn(U_{ba;ce})$, as $\Rn(U_{ba;ce})$ cannot exceed $d_e^2$.

In the particular case in which $d_e=d_a=2$, a one-qubit channel modeled using
a one-qubit mixed-state environment, there is a convenient parametrization of
the two-qubit unitary $U$ which allows one to show that the rank of its cross
operator can only take the values 1, 2, and 4; 3 is excluded.  See Sec.~IV of
\cite{Nlao03}, where this result is stated, in the notation employed there, as
Proposition 2.  This means that $\kp$ cannot be 3, because either $\sg=1$, in
which case $\kp\leq 2$ by \eqref{eqn40}, or else $\sg=2$, which means the
operator $S_{ce;cg}$ is nonsingular and the Kraus rank is equal to that of the
cross operator for $U$.  As noted above, the absence of $\kp=3$ was conjectured
on the basis of numerical work in \cite{Trao99}; so far as we know, ours is the
first analytic proof of this result.

In the case of a unitaries acting on two qutrits, $d_e=d_a=3$, it has been
shown \cite{Tysn03} that there are no restrictions on the rank of the cross
operator, which can take any value between 1 and 9, contrary to a conjecture in
\cite{Nlao03}.  Studies by one of us \cite{YuLi05} confirm that $d_e=d_a=2$ is
in this respect somewhat exceptional.

	\section{Summary and Open Questions}
\label{sct6}

The system of diagrams defined in Sec.~\ref{sct2} has been carefully
constructed to make it easy to convert quantum circuits into atemporal
diagrams, as illustrated in the various examples in Figs.~\ref{fgr6},
\ref{fgr7}, and \ref{fgr10}.  Nonetheless, it should be emphasized that the
arrows in atemporal diagrams have an abstract significance not connected with
the direction of time, and thus one need not be concerned as to whether
complicated diagrams, such as those in Fig.~\ref{fgr11} with arrows pointing in
various different directions, can be given a temporal interpretation.  One of
the principal advantages of map-state duality resides precisely in the
possibility of removing temporal references.  It is, for example, useful when
interpreting the diagram in Fig.~\ref{fgr7}(c) to remember that one can think
of $M_j$ as a quantum channel from $\HS_c$ to $\HS_b$ without requiring that
$\HS_b$ be at a time \emph{later} than the one at which Alice's measurement
takes place.

To be sure, the reader may well object that the atemporal feature of such
diagrams, however useful it may be for solving formal problems, serves to empty
them of any physical or intuitive content.  In response, note that such
diagrams are best thought of as pre-probabilities in the terminology of
\cite{Grff02}---useful for computing probabilities once appropriate
frameworks of quantum histories have been introduced---and not physical
reality (i.e., actual quantum histories).  Wave functions obtained by unitary
time development and used to calculate Born probabilities are also
pre-probabilities (Sec.~9.4 of \cite{Grff02}), and thus atemporal diagrams are
no less ``real'' than a variety of other tools used by quantum
physicists.  This response, while formally correct, conceals an interesting
question: when can one find a framework of quantum histories in which a given
atemporal diagram makes sense (i.e., assigns physically meaningful
probabilities) in terms of a narrative of Alice preparing something, Bob
measuring something, and the like?  Alice and Bob live in a world which is
irreversible in the thermodynamic sense, which makes preparation very different
from measurement, whereas microscopic quantum theory is reversible.  Connecting
the two is not an altogether trivial task, and while physicists typically
assign it to philosophers, there may be some aspects which physicists
themselves will have to disentangle in order to attain clear ideas about
quantum information.

We have given two applications of atemporal diagrams yielding results which
represent a generalization or clarification of work in the previous
literature. The first concerns unambiguous teleportation, for which our
discussion in Sec.~\ref{sct4}, in particular \ref{sct4c}, includes in a unified
scheme all the various examples and results in the previous literature for the
situation in which the three relevant Hilbert spaces have the same dimension
$d$.  By using the notion of the inverse of the shared entangled ket it is easy
to produce a variety of optimal protocols and understand how they are related
to each other.  (The corresponding problem of unambiguous dense coding, the
topic of a later paper by some of us \cite{Wuao05}, turns out to be much more
difficult to analyze.)  Of course, when teleportation is discussed outside the
``unambiguous'' framework by considering a mixed state as a resource, or
allowing for transmission of states with fidelity less than one, the problems
become more difficult, and it is not known whether or not atemporal diagrams
will aid in their solution.

The second application, showing that the Kraus rank of a one qubit noisy
channel with one-qubit mixed-state environment cannot have Kraus rank 3,
Sec.~\ref{sct5c}, while of somewhat limited interest in itself --- good
numerical evidence for it was available some time ago \cite{Trao99} --- raises
interesting questions about the relationship of a noisy channel (i.e., the
corresponding superoperator) to the ``cross operator'' of the isometry $V$, or
unitary $T$, used to model it, Fig.~\ref{fgr10}(b).  This is related to, but
not the same as the question of how much entanglement can be produced by a
unitary, or some more general operation, acting on a bipartite system; see the
extensive discussion in \cite{Nlao03}.  Since a unitary acting on two qubits
seems to be exceptional, \cite{Tysn03,YuLi05}, the Kraus rank does not look
like it will be useful for understanding what is special about other noisy
quantum channels based on a mixed-state environment. However, other properties
of the cross operator may be relevant.  A better understanding of the
connection between a quantum channel and the cross operator of the isometry, or
the properties of the dynamical operator (matrix), remains an open question
whose significance has, we believe, been somewhat sharpened by our diagrammatic
approach, even though it did not originate there.

	\section*{Acknowledgments}
We thank Y. Sun for comments on unambiguous discrimination, V. Gheorghiu for a
suggestion used in App.~\ref{sctpb}, and K. \.Zyczkowski for comments on the
bibliography.  The research described here received support from the National
Science Foundation through Grants PHY-0139974 and PHY-0456951.

\appendix
\numberwithin{equation}{section}

	\section{Appendix. Probability of an Unambiguous Operation}
\label{sctpa}

Let $K$ be a linear map from $\HS_a$ to $\HS_b$, where these are any two
Hilbert spaces.  We are interested in the probability that $K$ can be
carried out physically as an \emph{unambiguous operation} in the sense that one
could, at least in principle, construct an apparatus such that for any input
state $|\al\rgl\in\HS_a$, the output is guaranteed to be $K|\al\rgl\in\HS_b$,
provided an \emph{auxiliary system} at the end of the process has a property
$S$ indicating that the operation has been successful.

To be more precise, suppose there is a unitary transformation $T$ that maps
$\HS_a\ot\HS_x$ to $\HS_b\ot\HS_y$, where $\HS_x$ and $\HS_y$ are the Hilbert
spaces of the initial and final auxiliary systems (which could be the same if
$d_a=d_b$).  Assume the initial state is
$|\al\rgl\ot|\xi_0\rgl\in\HS_a\ot\HS_x$, with $|\xi_0\rgl$ fixed, and that
success corresponds to a subspace of $\HS_y$ with projector $S\in\hat\HS_y$
chosen in such a way that
\begin{equation}
   SV|\al\rgl = K|\al\rgl\ot |\eta_\al\rgl,
\label{Aeqn1}
\end{equation}
where  the isometry $V:\HS_a\ra\HS_{by}$ is defined by
\begin{equation}
  V|\al\rgl= T\Blp |\al\rgl\ot|\xi_0\rgl\Brp,
\label{Aeqn2}
\end{equation}
and $|\eta_\al\rgl\in\HS_y$ might depend on the initial $|\al\rgl$, as
suggested by the subscript.  Note that $K$, $T$, $|\xi_0\rgl$, $V$, and $S$ are
all regarded as fixed quantities characteristic of the apparatus, whereas the
initial $|\al\rgl$ is thought of as variable, or unknown, and the fixed
apparatus must perform in such a way that \eqref{Aeqn1} is true for \emph{any}
initial $|\al\rgl$.

None of the kets in \eqref{Aeqn1} need be normalized, and $K$ could be replaced
by $cK$, $c$ any nonzero constant, without altering the following discussion
(e.g., the constant could be absorbed in $|\eta_\al\rgl$).  That is, the
operation should be thought of as mapping rays (one-dimensional subspaces) in
$\HS_a$ to rays in $\HS_b$.  The key point is that in the case of success the
outcome, \eqref{Aeqn1},  must be a \emph{product} state; if it is entangled,
the desired result will arise with, at best, some probability less than 1.
Note also that \eqref{Aeqn1} is assumed to hold for \emph{all} $|\al\rgl$ in
$\HS_a$. If producing (the ray corresponding to) $K|\al\rgl$ for a
\emph{single} $|\al\rgl$ is all that is required, it is trivial to construct an
apparatus to do this with certainty.
One can interpret success as the positive outcome of an ideal measurement
carried out on $\HS_y$ to distinguish $S$ from $I-S$; e.g., a green light
flashes in case $S$ and a red light for $I-S$. Since we have allowed an
arbitrary auxiliary system, there is no need to talk about POVMs; projective
measurements are sufficient.  And if one uses a framework in which idealized
measurements indicate pre-existing properties (Ch.~17 of \cite{Grff02}), there
is no need to even talk about measurements.

The linearity of the operators in \eqref{Aeqn1} and the fact that it holds for
all $|\al\rgl$ in $\HS_a$ means---set $E=K$ and $F=SV$ in
App.~\ref{sctpb}---there is a \emph{fixed} ket $|\bar\eta\rgl\in\HS_y$,
independent of $|\al\rgl$, such that
\begin{equation}
  SV|\al\rgl = K|\al\rgl\ot |\bar\eta\rgl = \bar K|\al\rgl\ot |\eta_0\rgl,
\label{Aeqn3}
\end{equation}
where the second equality introduces the normalized quantities
\begin{equation}
  |\eta_0\rgl = |\bar\eta\rgl/\|\bar\eta\|,\quad \bar K = \|\bar\eta\| K,
\label{Aeqn4}
\end{equation}
with $\|\bar\eta\|=\sqrt{\lgl\bar\eta|\bar\eta\rgl}$ the usual norm.

The probability of success given an initial normalized state $|\al\rgl$,
$\|\al\|=1$,
is given by the Born rule
\begin{equation}
  \Pr(S\vb\al) = \lgl\al|V\ad SV|\al\rgl 
  = \lgl\al|\bar K\ad\bar K|\al\rgl\leq 1,
\label{Aeqn5}
\end{equation}
where the final inequality, necessarily true of a probability, can be checked
by writing $I_a=V\ad V$ as a sum of the two positive operators $V\ad SV$ and
$V\ad(I-S)V$. As this inequality holds for any (normalized) $|\al\rgl$, the
maximum eigenvalue of the positive operator $\bar K\ad\bar K$ cannot exceed 1.
Thus the probability of success is bounded by
\begin{equation}
  \Pr(S\vb\al)\leq \lgl\al|\hat K\ad\hat K|\al\rgl,
\label{Aeqn6}
\end{equation}
where $\hat K$ is defined as $cK$ when $c>0$ chosen so that the maximum
eigenvalue of $\hat K\ad\hat K$ is precisely 1.

In fact, the upper bound is achievable using the following strategy. Since
$I-\hat K\ad \hat K$ is a positive operator, the same is true of its positive
square root $\hat L$, and we can define an isometry
\begin{equation}
  V|\al\rgl = \hat K |\al\rgl\ot|\eta_0\rgl + \hat L |\al\rgl\ot|\eta_1\rgl,
\label{Aeqn7}
\end{equation}
using any two orthonormal states $|\eta_0\rgl$ and $|\eta_1\rgl$ in
$\HS_y$. With $S=|\eta_0\rgl\lgl\eta_0|$, \eqref{Aeqn3} holds with $\bar K=\hat
K$, making \eqref{Aeqn6} an equality.  (Extending $V$ to a unitary $T$ is a
straightforward exercise.)

	\section{Appendix. Vector constant of proportionality}
\label{sctpb}

Let $\VS_a$,  $\VS_b$ and  $\VS_c$  be any three linear spaces---we are
interested in Hilbert spaces, but the inner product plays no role in the
following argument---and suppose that $E:\VS_a\ra\VS_b$ and
$F:\VS_a\ra\VS_b\ot\VS_c$ are two linear maps such that
\begin{equation}
  F |\al\rgl = (E|\al\rgl)\ot |\gm_\al\rgl
\label{Beqn1}
\end{equation}
for every $|\al\rgl$ in  $\VS_a$, where $|\gm_\al\rgl$ might depend upon
$|\al\rgl$, as indicated by the subscript.

\emph{Theorem.}
Equation \eqref{Beqn1} implies that there is a fixed vector
$|\bar\gm\rgl\in\VS_c$ such that
\begin{equation}
  F |\al\rgl = (E|\al\rgl)\ot |\bar\gm\rgl.
\label{Beqn2}
\end{equation}

Even in the case in which $\VS_c$ is one-dimensional, the collection of
scalars, the result is not trivial, and it is useful to state it as:

\emph{Corollary.}
If $G$ and $E$ are two linear maps from $\VS_a$ to $\VS_b$ such that
\begin{equation}
  G|\al\rgl = c(\al) E |\al\rgl
\label{Beqn3}
\end{equation}
for all $|\al\rgl$, where $c(\al)$ is a scalar that may depend on $|\al\rgl$,
then there is a fixed scalar $\bar c$ such that $G=\bar c E$.

In constructing the proof, note that if $E|\al\rgl$, and hence $F|\al\rgl$, is
zero, $|\gm_\al\rgl$ can be anything.  Thus our the task is to show that for
all nonzero $E|\al\rgl$ the corresponding $|\gm_\al\rgl$ does not, in fact,
depend on $|\al\rgl$. Then if we set $|\bar\gm\rgl$ equal to this constant
vector, \eqref{Beqn2} will hold in all cases, including those in which
$E|\al\rgl$ vanishes.  Let $|\al_0\rgl$ and $|\al_1\rgl$ be any two distinct
elements of $\VS_a$ for which
\begin{equation}
  |\bt_0\rgl=E|\al_0\rgl\neq 0,\quad |\bt_1\rgl=E|\al_1\rgl\neq 0,
\label{Beqn4}
\end{equation}
and let $|\gm_0\rgl$ and $|\gm_1\rgl$ be the corresponding elements of $\VS_c$
in \eqref{Beqn1}.  In showing that $|\gm_1\rgl=|\gm_0\rgl$ we shall consider
three cases that exhaust all the possibilities.

Case 1. The vectors $|\al_0\rgl$ and $|\al_1\rgl$ are linearly dependent, say
$|\al_1\rgl=c|\al_0\rgl$. Then by linearity $|\bt_1\rgl=c|\bt_0\rgl$,
$F|\al_1\rgl=cF|\al_0\rgl$, and from \eqref{Beqn1} it follows at once that
$|\gm_0\rgl=|\gm_1\rgl$.

Case 2. The vectors $|\al_0\rgl$ and $|\al_1\rgl$ are linearly independent,
but $|\bt_0\rgl$ and $|\bt_1\rgl$ are linearly dependent, so that
\begin{equation}
  |\bt_1\rgl = c |\bt_0\rgl
\label{Beqn5}
\end{equation}
for some $c\neq 0$. This means that $E(|\al_1\rgl - c|\al_0\rgl)=0$
and therefore $F(|\al_1\rgl - c|\al_0\rgl)=0$.  Rewrite the second equality as
\begin{equation}
  F|\al_1\rgl = |\bt_1\rgl \ot|\gm_1\rgl = c F|\al_0\rgl
  = c|\bt_0\rgl\ot|\gm_0\rgl.
\label{Beqn6}
\end{equation}
Using \eqref{Beqn5} and $|\bt_1\rgl\neq 0$, see \eqref{Beqn4}, we conclude that
$|\gm_1\rgl = |\gm_0\rgl$.

Case 3. The vectors $|\bt_0\rgl$ and $|\bt_1\rgl$ are linearly independent.
Then use \eqref{Beqn1}, with $|\gm_\al\rgl=|\gm_2\rgl$ when
$|\al\rgl=|\al_0\rgl+|\al_1\rgl$, and linearity, to obtain
\begin{equation}
  F\Blp |\al_0\rgl+|\al_1\rgl\Brp  = 
  \Blp |\bt_0\rgl + |\bt_1\rgl\Brp \ot |\gm_2\rgl
  = |\bt_0\rgl\ot|\gm_0\rgl + |\bt_1\rgl\ot|\gm_1\rgl.
\label{Beqn7}
\end{equation}
Rewrite the second equality as
\begin{equation}
  |\bt_0\rgl\ot\Blp |\gm_2\rgl-|\gm_0\rgl\Brp  =  
  |\bt_1\rgl\ot\Blp |\gm_1\rgl-|\gm_2\rgl\Brp 
\label{Beqn8}
\end{equation}
to make it obvious that, since the $|\bt_j\rgl$ are linearly independent,
$|\gm_2\rgl=|\gm_0\rgl$ and $|\gm_1\rgl=|\gm_2\rgl$, so
$|\gm_1\rgl=|\gm_0\rgl$.


\end{document}